\documentclass[preprint,aps,eqsecnum,superscriptaddress]{revtex4}
\usepackage{graphicx,amsfonts}
\usepackage{amsmath}

\newcommand{\pa}{\partial}

\newcommand{\dk}{\frac{d^{4}k}{\left(2\pi\right)^{4}}}

\begin{document}

\title {\Large Superfield covariant analysis of the divergence structure
of noncommutative supersymmetric QED$_4$}

\author{A. F. Ferrari}
\author{H. O. Girotti}
\affiliation{Instituto de F\'{\i}sica, Universidade Federal do Rio Grande
do Sul, Caixa Postal 15051, 91501-970 - Porto Alegre, RS, Brazil}
\email{alysson, hgirotti, aribeiro@if.ufrgs.br}

\author{M. Gomes}
\author {A. Yu. Petrov}
\altaffiliation[Also at]{ Department of Theoretical Physics,
Tomsk State Pedagogical University
Tomsk 634041, Russia
(email: petrov@tspu.edu.ru)}
\affiliation{Instituto de F\'{\i}sica, Universidade de S\~{a}o Paulo,
 Caixa Postal 66318, 05315-970, S\~{a}o Paulo - SP, Brazil}
\email{mgomes, petrov, rivelles, ajsilva@fma.if.usp.br}

\author{A. A. Ribeiro}
\affiliation{Instituto de F\'{\i}sica, Universidade Federal do Rio Grande
do Sul, Caixa Postal 15051, 91501-970 - Porto Alegre, RS, Brazil}

\author{V. O. Rivelles}
\altaffiliation[Formerly at]{ Center for Theoretical Physics,
Massachusetts Institute of Technology,
Cambridge, MA 02139-4307, USA
(email: rivelles@lns.mit.edu)}
\author{A. J. da Silva}
\affiliation{Instituto de F\'{\i}sica, Universidade de S\~{a}o Paulo,
 Caixa Postal 66318, 05315-970, S\~{a}o Paulo - SP, Brazil}

\begin{abstract}
Commutative supersymmetric Yang-Mills is known to be renormalizable for
${\cal N} = 1, 2$, while finite for ${\cal N} = 4$. However, in the
noncommutative version of the model (NCSQED$_4$) the UV/IR mechanism
gives rise to infrared divergences which may spoil the perturbative
expansion.  In this work we pursue the study of the consistency of
NCSQED$_4$ by working systematically within the covariant superfield
formulation.  In the Landau gauge, it has already been shown for
${\cal N} = 1$ that the gauge field two-point function is free of
harmful UV/IR infrared singularities, in the one-loop
approximation. Here we show that this result holds without
restrictions on the number of allowed supersymmetries and for any
arbitrary covariant gauge.  We also investigate the divergence
structure of the gauge field three-point function in the one-loop
approximation. It is first proved that the cancellation of the leading
UV/IR infrared divergences is a gauge invariant
statement. Surprisingly, we have also found that there exist
subleading harmful UV/IR infrared singularities whose cancellation
only takes place in a particular covariant gauge. Thus, we conclude
that these last mentioned singularities are in the gauge sector and,
therefore, do not jeopardize the perturbative expansion and/or the
renormalization of the theory.
\end{abstract}

\maketitle
\newpage

\section{Introduction}
\label{sec:level1}

During last years noncommutative (NC) field theories have been
intensively studied. These theories emerged as the low energy limit of
the open superstring in the presence of an external magnetic field
($B$-field) \cite{SW} although nowadays they are interesting in their
own right (for a review see \cite{Nekr,Szabo,Girotti11}).

The most striking property of noncommutative field theories is
undoubtly the UV/IR mechanism, through which the ultraviolet
divergences (UV) are partly converted into infrared (IR) ones
\cite{Minw,Mat,RR1}. These infrared divergences \cite{footnote1} may
be so severe that the perturbative expansion of the theory becomes
meaningless. Hence, the key point about the consistency of a
noncommutative field theory is whether these divergences cancel out.

So far, only one four-dimensional noncommutative theory is known to be
renormalizable, the Wess-Zumino model \cite{Girotti1,Buchbinder1}. In
this case supersymmetry plays an essential role because it improves
the ultraviolet behavior and, therefore, the UV/IR mechanism only
generates mild UV/IR infrared divergences which do not spoil the
renormalization program. In three space-time dimensions we are aware
of at least two noncommutative renormalizable models: the
supersymmetric $O(N)$ nonlinear sigma model \cite{sig} and the $O(N)$
supersymmetric linear sigma model in the limit $N \rightarrow \infty$
\cite{linsm}. 

As for nonsupersymmetric gauge theories, the UV/IR
mechanism breaks down the perturbative approach
\cite{Mat,RR1,Hay,SJ,Armoni,Bonora,FL,Gur,Nichol}. Nevertheless, we can entertain
the hope that noncommutative supersymmetric gauge theories are free from nonintegrable UV/IR infrared singularities and, furthermore, renormalizable. We are aware of the following
results concerning noncommutative supersymmetric gauge field theories:  

1) By working with the formalism of component fields \cite{Mat,RR1} it
has been shown that the dangerous UV/IR infrared divergences cancel in the
one-loop contributions to the gauge field two and three-point
functions. The two-point function turns out to contain quadratic and
logarithmic UV divergences. Dimensional regularization takes care of
the first ones while the last ones must be renormalized. As for the
harmful infrared divergences originating through the UV/IR mechanism,
they are only quadratic and cancel out within a supersymmetric
multiplet. The three-point function is linearly UV divergent by power
counting. However, this time, the leading UV divergences vanish by
symmetric integration, while the IR poles originating from them cancel
out among themselves.

2) By using the superfield formalism Bichl et al. \cite{bichl} calculated, in the Landau gauge and in the absence of matter (${\cal N} = 1$), the one-loop contributions to the two-point function of the gauge superfield. Only quadratic and logarithmic UV divergences are present and one deals with them as indicated in the previous paragraph. The quadratic infrared poles in the nonplanar part of the amplitude again cancel while the linear ones do not arise. The superfield formulation represents an improvement with respect to the component field formulation because supersymmetry is explicitly preserved at all stages of the calculation.  

3) Zanon and collaborators \cite{Za1,Zanon} used the background field
method to evaluate the one-loop contributions to the field strength two-point
functions in ${\cal N} = 1, 2$ supersymmetric Yang-Mills theories, where
only logarithmic divergences were found. The three-point function was
shown to vanish. For ${\cal N}=4$ they demonstrated that, up to one loop, there
are no divergences at all.

This paper is dedicated to pursue further the study, within the
superfield formulation, of the consistency of NCSQED$_4$ in an
arbitrary covariant gauge. We analyze the divergence structure induced
by the UV/IR mechanism in the two and three-point gauge field Green functions.

In Section 2 we establish our definitions and conventions and present
the gauge invariant action describing the dynamics of NCSQED$_4$ in
${\cal N} = 1$ superspace. Next, the gauge fixing and the
Faddeev-Popov terms are found. Finally, we add chiral matter
superfields and derive the Feynman rules of NCSQED$_4$ with extended
supersymmetry.

We start, in Section 3, by reviewing the cancellation of the leading
UV/IR infrared divergences in the one-loop corrections to the
two-point function of the gauge superfield\cite{bichl}. A
straightforward generalization shows that these results also hold for
extended supersymmetry and/or when the theory is formulated in an
arbitrary covariant gauge.

In Section 4 we compute the one-loop corrections to the three-point
functions of the gauge superfield in an arbitrary covariant
gauge. This is done for ${\cal N} = 1, 2, 4$. From power counting
follows that the amplitude is at the most quadratically divergent. As
far as the planar part is concerned, dimensional regularization takes
care of the quadratic UV divergences, the linear ones vanish by
symmetric integration, while the logarithmic divergences are to be
eliminated through renormalization. As for the nonplanar part, the
UV/IR mechanism will be seen not to give rise to quadratic IR
divergences but only to linear and logarithmic ones. Interestingly
enough, the linear IR divergences arise from two different sources: a)
integrals which, by power counting, are quadratically UV divergent but
whose Moyal phase factor not only regularizes them but also lowers the
degree of the IR divergence, b) integrals which are linearly UV
divergent by power counting but regularized by the
noncommutativity. The softening mechanism mentioned in a) also
contributes IR logarithmic divergences, which, nevertheless, do not
jeopardize the perturbative expansion.

The conclusions are contained in Section 5.

\section{The action and Feynman rules for NCSQED$_4$}
\label{sec:level2}

\subsection{The action}

In ${\cal N} = 1$ superspace NCSQED$_4$ is described by the
nonpolynomial action \cite{SGRS,footnoteSGRS}

\begin{equation}
\label{II-1}
 S_V\,=\,-\,\frac{1}{2g^{2}}\int d^{8}z\left(e^{-gV}*D^{\alpha }e^{gV}\right)*\overline{D}^{2}\left(e^{-gV}*D_{\alpha }e^{gV}\right)\,,
\end{equation}

\noindent
where $g$ is the coupling constant, $V$ is a real vector gauge superfield, 

\begin{equation}
\label{II-2}
e^{gV}\,=\,\sum _{n\geq 0}\frac{1}{n!}\left(g V \right)^{*n}\,\equiv\,\sum _{n\geq 0}\frac{1}{n!}\,\left(
\begin{array}{c}
\underbrace{gV * gV* ...gV}\\
n-times
\end{array}\right)\,,
\end{equation}

\noindent
and $*$ denotes Moyal product of operators, i.e.,

\begin{eqnarray}
\label{II-3}
&&\phi_1(x) \ast \phi_2(x)\,=\,\phi_1(x)\,exp \left(\frac{i}{2} \,\overleftarrow{\frac{\pa}{\pa x ^{\mu}}}\,\Theta^{\mu \nu}\,\overrightarrow{\frac{\pa}{\pa x^{\nu}}}\right)\,\phi_2(x)\nonumber\\
&&=\,\sum_{n = 0}^{\infty}\,\left(\frac{i}{2}\right)^n\,\frac{1}{n!}\,\left[\pa_{\mu_1} \pa_{\mu_2}...\pa_{\mu_n}\,\phi_1(x)\right]\,\Theta^{\mu_1 \nu_1} \Theta^{\mu_2 \nu_2} ... \Theta^{\mu_n \nu_n}\,\left[\pa_{\nu_1} \pa_{\nu_2}...\pa_{\nu_n}\,\phi_2(x)\right]\,.
\end{eqnarray}

\noindent
Here, $\Theta^{\mu \nu}$ is the antisymmetric real constant matrix characterizing the noncommutativity of the underlying space-time. The expression

\begin{eqnarray}
\label{II-301}
&&\int d^{4}x\varphi _{1}\left(x\right)*\cdots *\varphi _{n}\left(x\right)\nonumber\\
&&=\,\int \left(\prod _{j=1}^{n}\frac{d^{4}k_{j}}{\left(2\pi \right)^{4}}\right)\left(2\pi \right)^{4}\delta \left(\sum_{1}^{n} k_{j}\right)e^{-i\sum _{i<j}k_{i}\wedge k_{j}}\varphi _{1}\left(k_{1}\right)\cdots \varphi _{n}\left(k_{n}\right)\,,
\end{eqnarray}

\noindent
where

\begin{equation}
\label{II-302}
k_i \wedge k_j\,=\,\frac{1}{2}\,k^{\mu}_i\,k^{\nu}_j\,\Theta_{\mu \nu}\,,
\end{equation}

\noindent
will play a relevant role for determining the Feynman rules in the theory.

Under the group $U(1)$ of gauge transformations

\begin{equation}
\label{II-4}
U\,=\,e^{ig\Lambda}\,=\,\sum _{n\geq 0}\frac{1}{n!}\left(ig\Lambda \right)^{*n}\,,
\end{equation}
  
\noindent
with $\Lambda$ (${\bar \Lambda} = \Lambda^{\dagger}$) a chiral (antichiral) superfield, $V$ transforms as follows

\begin{equation}
\label{II-5}
e^{gV}\rightarrow e^{-ig {\Lambda }}*e^{gV}*e^{ig \overline{\Lambda} }\,,
\end{equation}

\noindent
thus leaving $S$ invariant.

In future, we shall be needing the expansion of $S$ in powers of $g$, up to the order $g^3$. To this end we first recall the identity \cite{Buchbinder2}

\begin{eqnarray}
\label{II-6}
&&e^{-gV}*D_{\alpha }e^{gV}\,\nonumber\\&=&\,gD_{\alpha }V-\frac{g^{2}}{2!}\left[V,D_{\alpha }V\right]_{*}+\frac{g^{3}}{3!}\left[V,\left[V,D_{\alpha }V\right]_{*}\right]_{*}-\frac{g^{4}}{4!}\left[V,\left[V,\left[V,D_{\alpha }V\right]_{*}\right]_{*}\right]_{*}\nonumber\\
&&+\,\frac{g^5}{5!}\,\left[V,\left[V,\left[V, \left[V, D_{\alpha }V\right]_{*}\right]_{*}\right]_{*}\right]_{*}\,+\,\cdots\,.
\end{eqnarray}

\noindent
Then, after by part integrations and by exploring the properties of
the Moyal product \cite{footnote2} one obtains

\begin{equation}
\label{II-7}
S_V\,=\,S_V^{(0)}\,+\,g\,S_V^{(1)}\,+\,g^2\,S_V^{(2)}\,+\,g^3\,S_V^{(3)}\,+\,\cdots\,,
\end{equation}

\noindent
where

\begin{equation}
\label{II-8}
S_V^{(0)}\,=\,\frac{1}{2}\int d^{8}z\,V\,D^{\alpha }\overline{D}^{2}D_{\alpha }V\,,
\end{equation}

\begin{equation}
\label{II-9}
S_V^{(1)}=\frac{1}{2}\int d^{8}z\,\overline{D}^{2}D^{\alpha }V*\left[V,D_{\alpha }V\right]_{*}\,,
\end{equation}

\begin{equation}
\label{II-10}
S_V^{(2)}=-\,\int d^{8}z\left\{ \frac{1}{8}\,\left[V,D^{\alpha }V\right]_{*}*\overline{D}^{2}\left[V,D_{\alpha }V\right]_{*}+\frac{1}{6}\,\overline{D}^{2}D^{\alpha }V*\left[V,\left[V,D_{\alpha }V\right]_{*}\right]_{*}\right\} \,,
\end{equation}

\begin{eqnarray}
\label{II-11}
S_V^{(3)}&=&\frac{1}{12}\int d^{8}z\left\{ \frac{1}{2}\,\overline{D}^{2}D^{\alpha }V*\left[V,\left[V,\left[V,D_{\alpha }V\right]_{*}\right]_{*}\right]_{*}\right.\nonumber\\
 &&\left.+\left[V,\left[V,D^{\alpha }V\right]_{*}\right]_{*}*\overline{D}^{2}\left[V,D_{\alpha }V\right]_{*}\right\}\,.
\end{eqnarray}

As usual, gauge fixing is implemented by adding to the action $S_V$ the covariant term

\begin{equation}
\label{II-13}
S_{gf}\,=\,-\frac{a}{2}\int d^{8}zV\left\{ D^{2},\overline{D}^{2}\right\} V\,,
\end{equation}

\noindent
where $a$ is a real number labeling the gauge. Clearly, 

\begin{equation}
\label{II-14}
S_V^{(0)}\,+\,S_{gf}  =  \frac{1}{2}\int d^{8}z\,V\left(\Box +\left(1-a\right)\left\{ D^{2},\overline{D}^{2}\right\} \right)V\,,
\end{equation}

\noindent
as seen from Eqs. (\ref{II-8}) and (\ref{II-13}). 

For the covariant gauge $a$, the Faddeev-Popov determinant reads

\begin{equation}
\label{II-16}
\Delta ^{-1}\left[V\right] \, = \,\int \mathcal{D}c\,\mathcal{D}c^{\prime }\,\mathcal{D}\overline{c}\,\mathcal{D}\overline{c}^{\prime }\,e^{-\,\int d^{8}z\,\left[c\left(z\right)+\overline{c}\left(z\right)\right]\,\delta V\left(z\right)}|_{\Lambda =c^{\prime }\, ;\, \overline{\Lambda }=\overline{c}^{\prime }}\,.
\end{equation}

\noindent
Here $c, \overline{c}= c^{\dagger}, c^{\prime}, \overline{c}^{\prime}={c^{\prime}}^{\dagger}$ are the ghost fields while $\delta V$ denotes the change in $V$ provoked by an infinitesimal gauge transformation. One readily obtains from (\ref{II-5}) that

\begin{equation}
\label{II-17}
\delta V \, = \, iL_{\frac{g}{2}V}\,\left[-\left(\Lambda +\overline{\Lambda }\right)+\left(\coth L_{\frac{g}{2}V}\right)\left[\overline{\Lambda }-\Lambda \right]\right]\,,
\end{equation}

\noindent
where

\begin{equation}
\label{II-18}
L_{A}\left[B\right]\equiv \left[A,B\right]_{*}\,.
\end{equation}

\noindent
After recalling the Laurent expansion of $\coth x$, around $x = 0$, one 
arrives at

\begin{eqnarray}
\label{II-19}
\delta V & = & iL_{\frac{g}{2}V}\left[-\left(\Lambda +\overline{\Lambda }\right)\right]
\nonumber\\
&+&iL_{\frac{g}{2}V}\left[L_{\frac{g}{2}V}^{-1}\left[\overline{\Lambda }-\Lambda \right]+\frac{1}{3}L_{\frac{g}{2}V}\left[\overline{\Lambda }-\Lambda \right] \,+\,\cdots\right]\nonumber \\
 & = & i\left(\overline{\Lambda }-\Lambda \right)-\frac{ig}{2}\left[V,\Lambda +\overline{\Lambda }\right]_{*}+\frac{ig^{2}}{12}\left[V,\left[V,\overline{\Lambda }-\Lambda \right]_{*}\right]_{*}
\,+\,\cdots\,.
\end{eqnarray}

\noindent
Therefore, by going back with Eq. (\ref{II-19}) into Eq. (\ref{II-16})
one finds for the ghost action the following expression

\begin{equation}
\label{II-191}
S_{gh}\,=-\,S_{gh}^{(0)}\,+\,g\,S_{gh}^{(1)}\,+\,g^2\,S_{gh}^{(2)}\,+\,\cdots\,,
\end{equation}

\noindent
where

\begin{equation}
\label{II-201}
S_{gh}^{(0)}\,=\,- \,\int d^{8}z\left(c+\overline{c}\right) \left(\overline{c}^{\prime }-c^{\prime }\right)\,,
\end{equation}

\begin{equation}
\label{II-202}
S_{gh}^{(1)}\,=\,\,\frac{1}{2}\int d^8z \,\left(c+\overline{c}\right)\left[V,c^{\prime }+\overline{c}^{\prime }\right]_{*}\,,
\end{equation}

\begin{equation}
\label{203}
S_{gh}^{(2)}\,=\,-\frac{1}{12}\int d^8z \,\left(c+\overline{c}\right)\left[V,\left[V, \overline{c}^{\prime }-c^{\prime }\right]_{*}\right]_{*}\,.
\end{equation}

In addition to the real vector superfield we introduce now a chiral
matter superfield $\Phi$ in the adjoint representation. This enables
us to construct a theory in which the ${\cal N} = 2$ supersymmetry is
realized. The generalization to ${\cal N} = 4$ is straightforward and
will be done afterwards. The corresponding action describing the free
matter superfield as well as its interaction with the gauge superfield
reads

\begin{equation}
\label{II-21}
S_{m}\,=\,\int d^{8}z\overline{\Phi }*e^{-gV}*\Phi *e^{gV}\,,
\end{equation}

\noindent
whose invariance under supergauge transformations follows from (\ref{II-5}) together with 

\begin{equation}
\label{II-21a}
\Phi\rightarrow \Phi^{\prime}\,=\,e^{ig {\bar \Lambda }}*\Phi*e^{ ig {\Lambda} }\,,\,\,\,\,\,\,\,\,\,\,\,\,{\bar \Phi}\rightarrow {\bar \Phi}^{\prime}\,=\,e^{- ig {\bar \Lambda }}*{\bar \Phi}*e^{-ig {\Lambda} }\,.
\end{equation}

The first four terms of the expansion of $S_m$ as a power series of $g$, 

\begin{equation}
\label{II-22}
S_m\,=\,S_m^{(0)}\,+\,g\,S_m^{(1)}\,+\,g^2\,S_m^{(2)}\,+\,g^3\,S_m^{(3)}\,+\,\cdots\,,
\end{equation}

\noindent
are found to be

\begin{equation}
\label{II-23}
S_m^{(0)}\,=\,\int\,d^8z\,\overline{\Phi} {\Phi}\,, 
\end{equation} 

\begin{equation}
\label{II-24}
S_m^{(1)}\,=\,-\int d^{8}z \,\overline{\Phi }*\left[V , \Phi\right]_{*}\,,
\end{equation}

\begin{equation}
\label{II-25}
S_m^{(2)}\,=\,\frac{1}{2}\int d^{8}z\,\overline{\Phi }*\left[V,\left[V,\Phi \right]_{*}\right]_{*}\,,
\end{equation}

\begin{equation}
\label{II-26}
S_m^{(3)}\,=\,-\frac{1}{6}\int d^{8}z\,\overline{\Phi }*\left[V,\left[V,\left[V,\Phi \right]_{*}\right]_{*}\right]_{*}\,.
\end{equation}

\subsection{Feynman rules}

From the quadratic part of the action $S_V^{(0)} + S_{gf} + S_{gh}^{(0)} + S_m^{(0)}$ one obtains, through standard manipulations, the free propagators 

\begin{subequations}
\label{II-28}
\begin{eqnarray}
&&\Delta_{VV}(z_1 - z_2)\,=\,\frac{i}{\Box}\,\left[ 1\,+\,( 1 - a)\,
\frac{1}{\Box}\,\{D_1^2 , {\overline D}_1^2 \} \right]\,\delta^8(z_1 - z_2)\,,\label{mlett:aII-28}\\
&&\Delta_{{\overline c} c^{\prime}}(z_1 - z_2)\,=\,-\,\frac{i}{\Box}\,D_1^2 \,\overline{D}_2^2\,\delta^8(z_1 - z_2) ,\label{mlett:bII-28}\\
&&\Delta_{c {\overline c}^{\prime}}(z_1 - z_2)\,=\,\frac{i}{\Box}\,{\overline D}_1^2 \,{D}_2^2\,\delta^8(z_1 - z_2) \,,\label{mlett:cII-28}\\
&&\Delta_{\Phi \overline{\Phi}}(z_1 - z_2)\,=\,-\,\frac{i}{\Box}\,{\overline D}_1^2 \,{D}_2^2\,\delta^8(z_1 - z_2)\,,\label{mlett:dII-28}
\end{eqnarray}
\end{subequations}

\noindent
corresponding to the gauge, ghosts and matter superfields,
respectively. They are depicted in Fig. {\ref{propagators}.

On the other hand, the interacting part of the total action together
with Eq. (\ref{II-301}) enable us to find the elementary vertices
$\Gamma^{(0)}$ in the theory. They are displayed in
Fig. \ref{vertices}. In an obvious notation

\begin{equation}
\label{II-29}
\Gamma_{(\overline{D}^2DV)(DV)V}^{(0)}(k_1, k_2, k_3)\,=\,g\,{\cal V}_{3}(k_1, k_2, k_3)\,,
\end{equation}

\begin{subequations}
\label{II-2901}
\begin{eqnarray}
&& \Gamma_{(\overline{D}^2DV)(DV)VV}^{(0)}(k_1, k_2, k_3, k_4)
\,=\,-\,\frac{ig^2}{12}\,{\cal V}^{(1)}_{4}(k_1, k_2, k_3, k_4)\,,\label{mlett:aII-2901}\\
&& \Gamma_{V(DV)(\overline{D}V)({\overline D}DV)}^{(0)}(k_1, k_2, k_3, k_4)
\,=\,ig^2\,{\cal V}^{(2)}_{4}(k_1, k_2, k_3, k_4)\,,\label{mlett:bII-2901}\\
&& \Gamma_{V(DV)(\overline{D}^2V)(DV)}^{(0)}(k_1, k_2, k_3, k_4)
\,=\,ig^2\,{\cal V}^{(2)}_{4}(k_1, k_2, k_3, k_4)\,,\label{mlett:cII-2901}
\end{eqnarray}
\end{subequations}

\begin{subequations}
\label{II-2902}
\begin{eqnarray}
&& \Gamma_{(\overline{D}^2DV)(DV)VVV}^{(0)}(k_1, k_2, k_3, k_4, k_5)\,=\,-\,\frac{g^3}{36}\,{\cal V}_{5}^{(1)}(k_1, k_2, k_3, k_4, k_5)\,,\label{mlett:aII-2902}\\
&& \Gamma_{ V V (DV) (\overline{D}DV)({\overline D} V)}^{(0)}(k_1, k_2, k_3, k_4, k_5)\,=\,-\,\frac{2g^3}{3}\,{\cal V}_{5}^{(2)}(k_1, k_2, k_3, k_4, k_5)\,,\label{mlett:bII-2902}\\
&& \Gamma_{ V V (DV) (\overline{D}^2 V)(DV)}^{(0)}(k_1, k_2, k_3, k_4, k_5)\,=\,\frac{ig^3}{12}\,{\cal V}_{5}^{(3)}(k_1, k_2, k_3, k_4, k_5)\,,\label{mlett:cII-2902}
\end{eqnarray}
\end{subequations}

\begin{subequations}
\label{II-2903}
\begin{eqnarray}
&&\Gamma_{{\overline c}^{\prime} V c}^{(0)}(k_1, k_2, k_3)\,=\,g\,{\cal V}_{3}(k_1, k_2, k_3)\,,\label{mlett:aII-2903}\\
&&\Gamma_{{c}^{\prime} V {\overline c}}^{(0)}(k_1, k_2, k_3)\,=\,g\,{\cal V}_{3}(k_1, k_2, k_3)\,,\label{mlett:bII-2903}\\
&&\Gamma_{{c}^{\prime} V c}^{(0)}(k_1, k_2, k_3)\,=\,-\,g\,{\cal V}_{3}(k_1, k_2, k_3)\,,\label{mlett:cII-2903}\\
&&\Gamma_{{\overline c}^{\prime} V {\overline c}}^{(0)}(k_1, k_2, k_3)\,=\,-\,g\,{\cal V}_{3}(k_1, k_2, k_3)\,\label{mlett:dII-2903}\,,
\end{eqnarray}
\end{subequations}

\begin{subequations}
\label{II-2904}
\begin{eqnarray}
&& \Gamma_{\overline{c}^{\prime}V V c }^{(0)}(k_1, k_2, k_3, k_4)\,=\,-\,\frac{i g^2}{6}\,{\cal V}^{(1)}_{4}(k_1, k_2, k_3, k_4)\,,\label{mlett:aII-2904}\\
&& \Gamma_{c^{\prime}V V {\overline c} }^{(0)}(k_1, k_2, k_3, k_4)\,=\,\frac{i g^2}{6}\,{\cal V}^{(1)}_{4}(k_1, k_2, k_3, k_4)\,,\label{mlett:bII-2904}\\
&& \Gamma_{c^{\prime}V V c }^{(0)}(k_1, k_2, k_3, k_4)\,=\,\frac{i g^2}{6}\,{\cal V}^{(1)}_{4}(k_1, k_2, k_3, k_4)\,,\label{mlett:cII-2904}\\
&& \Gamma_{\overline{c}^{\prime}V V {\overline c} }^{(0)}(k_1, k_2, k_3, k_4)\,=\,-\,\frac{i g^2}{6}\,{\cal V}^{(1)}_{4}(k_1, k_2, k_3, k_4)\,,\label{mlett:dII-2904}\,
\end{eqnarray}
\end{subequations}

\begin{subequations}
\label{II-2905}
\begin{eqnarray}
&& \Gamma_{\overline{\Phi} V \Phi}^{(0)}(k_1, k_2, k_3)\,
=\,2\,g\,{\cal V}_{3}(k_1, k_2, k_3)\,,\label{mlett:aII-2905}\\
&& \Gamma_{\overline{\Phi} V V \Phi}^{(0)}(k_1, k_2, k_3, k_4)\,
=\,i\,g^2\,{\cal V}_{4}^{(1)}(k_1, k_2, k_3, k_4)\,,\label{mlett:bII-2905}\\
&& \Gamma_{\overline{\Phi}V V V \Phi}^{(0)}(k_1, k_2, k_3, k_4, k_5)\,
=\,-\,\frac{g^3}{9}\,{\cal V}_{5}^{(1)}(k_1, k_2, k_3, k_4, k_5)\,.
\label{mlett:cII-2905}
\end{eqnarray}
\end{subequations}

\noindent
Here, 

\begin{subequations}
\label{II-30}
\begin{eqnarray}
&& {\cal V}_{3}(k_1, k_2, k_3)\,=\,\sin (k_1 \wedge k_2)\,,
\label{mlett:aII-30}\\
&& {\cal V}^{(1)}_{4}(k_1, k_2, k_3, k_4)\,=\,\cos (k_2 \wedge k_3)\,\cos (k_1 \wedge k_4)\,-\,\cos(k_1 \wedge k_2\,-\,k_3 \wedge k_4)\,,\label{mlett:bII-30}\\
&& {\cal V}^{(2)}_{4}(k_1, k_2, k_3, k_4)\,=\,\frac{1}{2}\,\left[\sin (k_1 \wedge k_2)\sin (k_3 \wedge k_4)-\sin (k_1 \wedge k_4)\sin (k_2 \wedge k_3)\right]\,,\label{mlett:cII-30}\\
&& {\cal V}^{(1)}_{5}(k_1, k_2, k_3, k_4, k_5)\,=\,\left[2\,\cos( k_4 \wedge k_5)\,\cos( k_3 \wedge k_4\,+\,k_3 \wedge k_5) \right.\nonumber\\
&&{\phantom{abc}}+ \left. \, \cos(- k_3 \wedge k_4\,+\,k_4 \wedge k_5\,+\,k_3 \wedge k_5)\,\right]\,\sin(k_1 \wedge k_2)\nonumber\\
&&{\phantom{abc}} + \,3\,\left[\cos(- k_2 \wedge k_3\,+\,k_2 \wedge k_5\,+\,k_3 \wedge k_5)\,\sin(k_1 \wedge k_4)\right.\nonumber\\
&& {\phantom{abc}}+ \left. \,\cos(- k_2 \wedge k_4\,+\,k_4 \wedge k_5\,+\,k_2 \wedge k_5)\,\sin(k_1 \wedge k_3)\right.\nonumber\\
&&{\phantom{abc}} + \left.\,\cos(- k_2 \wedge k_4\,-\,k_3 \wedge k_4\,+\,k_2 \wedge k_3)\,\sin(k_1 \wedge k_5)\right]\,,\label{mlett:dII-30}\,\\
&& {\cal V}^{(2)}_{5}(k_1, k_2, k_3,k_4,k_5)\,=\,2 \sin(p_1 \wedge p_3) \sin(p_2 \wedge p_3) \cos(p_1 \wedge p_2)  \nonumber\\
&& {\phantom{abc}}+ \sin(p_1 \wedge p_2)\,\left[ \sin( p_2 \wedge p_3 - p_1 \wedge p_3) 
\right]\,,\label{mlett:eII-30}\\
&& {\cal V}^{(3)}_{5}(k_1, k_2, k_3,k_4,k_5)\,=\,2\,i\,\sin(p_1 \wedge p_2)\cos(p_2 \wedge p_4)\cos(p_3 \wedge p_5)
\nonumber\\
&&{\phantom{abc}}+ \, \exp(-i p_1 \wedge p_2)\,\cos(p_3 \wedge p_4 + p_3 \wedge p_5 + p_4 \wedge p_5)\nonumber\\
&& {\phantom{abc}}+ \, \exp(-i p_1 \wedge p_4)\,\cos(p_3 \wedge p_2 + p_3 \wedge p_5 + p_2 \wedge p_5)\nonumber\\
&&{\phantom{abc}}- \, \exp(-i p_1 \wedge p_3)\,\cos(p_2 \wedge p_5 + p_2 \wedge p_4 - p_4 \wedge p_5)\nonumber\\
&&{\phantom{abc}}- \, \exp(-i p_1 \wedge p_5)\,\cos(p_2 \wedge p_3 + p_2 \wedge p_4 - p_3 \wedge p_4)\,,\label{mlett:fII-30}
\end{eqnarray}
\end{subequations}

\noindent
the momenta are taken positive when entering the vertex and momentum
conservation holds in all vertices.

We close this Section by pointing out that the superficial degree of
divergence of a generic Feynman graph ${\cal G}$ is given by
\cite{SGRS}

\begin{equation}
\label{II-32}
d[{\cal G}]\,=\,2\,-\,E_c\,,
\end{equation}

\noindent
where $E_c$ is the number of external chiral lines. As known, in a
noncommutative quantum field theory, a generic Feynman graph ${\cal
G}$ will decompose into planar and nonplanar parts. The superficial
degree of UV divergence of the planar part is measured by $d[{\cal
G}]$. The nonplanar part is free of UV divergences but afflicted by IR
singularities generated through the UV/IR
mechanism \cite{Minw,footnote1}, in this last connection $d[{\cal G}]$
also gives the highest possible degree of the IR divergences.

\section{One loop-contributions to the vector gauge superfield two-point function $\Gamma^{(1)}_{VV}$.}
\label{sec:level3}

The cancellation of the harmful UV/IR infrared divergences in
$\Gamma^{(1)}_{VV}$ was already proved in \cite{bichl} for ${\cal
N} = 1$ and by working in the Landau gauge. Here, the proof is
generalized by showing that the just mentioned cancellation takes
place for an arbitrary covariant gauge and extended supersymmetry.

Let us first concentrate on the graphs involving either a $V$ tadpole
or a $V$ loop (see Fig. \ref{twopoint}). Since there are no external
chiral lines, $d[{\cal G}] = 2$. Now, only those graphs with all $D$
factors in the internal lines may exhibit quadratic UV
divergences. Diagrams with a factor $D$ and/or a ${\overline D}$ on
the external lines can at the most be linearly divergent. Any other
combination of $D$'s on the external lines corresponds to
contributions which are logarithmically divergent or finite. These
follows from the $D$-algebra alone \cite{SGRS}. However, one is to take
into account also the noncommutativity, which gives origin to a
trigonometric factor that modifies the Feynman integrands. The
combination of these two ingredients rules out, for the diagrams under
analysis, the UV and UV/IR infrared linearly divergent terms. Hence,
in this case, only quadratic divergences may jeopardize the 
consistency of the theory. They are contained in graphs (a),  (b)  and
(c) in Fig. \ref{twopoint}.

From the Feynman rules derived in Section 2, we found that the
contribution ${\Gamma} _{VV;3{\rm a}}^{(1)}$ arising from the $V$ tadpole
diagram is given by

\begin{eqnarray}
\label{III-1}
&&{\Gamma}_{VV;3{\rm a}}^{(1)}(p)\nonumber\\
&=&-\frac{g^{2}}{6}\int \frac{d^{4}k}{\left(2\pi \right)^{4}}d^{4}
\theta_1 d^{4}\theta_2 \mathcal{V}^{(1)}_{4}\left(-k,p,-p,k\right)
\frac{\delta_{12}}{k^2}\left(\overline{D}_1^{2} 
D_1^{\alpha}D_{2 \alpha} \delta_{12} \right)V\left(p,\theta_1 \right)V
\left(-p,\theta_2 \right).
\end{eqnarray}

\noindent
Here, a factor $2$ coming from the permutation of the external legs
has already been taken into account. Moreover, we note that the term
proportional to $(1 -a)$ in the right hand side of
(\ref{mlett:aII-28}) does not contribute.

From (\ref{mlett:bII-30}) one finds that

\begin{equation}
\label{III-2}
\mathcal{V}_{ 4}^{(1)}\left(-k,p,-p,k\right)\,=\,2\sin ^{2}(k\wedge p)\,.
\end{equation}

\noindent
After $D$-algebra manipulations, one ends up with

\begin{equation}
\label{III-3}
\Gamma_{VV;3{\rm a}}^{(1)}(p)\,=\,\frac{2}{3}\,g^2\,A\,,
\end{equation}

\noindent
where

\begin{equation}
\label{III-4}
A\,\equiv\,\int \frac{d^{4}k}{\left(2\pi \right)^{4}}\,d^{4}\theta\, \frac{\sin ^{2}(k\wedge p)}{k^{2}}\,V\left(p,\theta \right)\,V\left(-p,\theta \right)\,.
\end{equation}

\noindent
The planar part of $A$ only contains quadratic UV divergences, while the nonplanar one only develops quadratic IR infrared singularities. 

The amplitudes associated with diagrams (b) and (c) of
Fig. \ref{twopoint} are, respectively,

\begin{eqnarray}
\label{III-5a}
\Gamma_{VV;3{\rm b}}^{(1)}(p)\,&=&\, 2\, \times\, \frac{1}{2}\,g^2\,\int \frac{d^4k}{(2 \pi)^4}\,d^4\theta_1\,d^4\theta_2\,{\cal V}_3(k - p, p, -k)\,{\cal V}_3(k, -p, -k + p)\,\nonumber\\
&\times&\left[-\,\frac{1}{k^2 (k + p)^2}\right]
D^{\alpha}_1\,{\overline D}^2_2\,D^{\beta}_2\,\delta_{12}\,D_{2 \beta}\, {\overline D}^2_1\,D_{1 \alpha}\,\delta_{12}\,V(p, \theta_1)\,V(- p, \theta_2)\,,
\end{eqnarray}

\begin{eqnarray}
\label{III-5b}
\Gamma_{VV;3{\rm c}}^{(1)}(p)\,&=&\, 2\, \times\, \frac{1}{2}\,g^2\,\int \frac{d^4k}{(2 \pi)^4}\,d^4\theta_1\,d^4\theta_2\,{\cal V}_3(k-p, p, -k)\,{\cal V}_3(- k+p, -p, k )\,\nonumber\\
&\times&\left[-\,\frac{1}{k^2 (k + p)^2}\right]
D^{\alpha}_1\,D^{\beta}_2\,\delta_{12}\,{\overline D}^2_2\,D_{2 \beta}\, {\overline D}^2_1\,D_{1 \alpha}\,\delta_{12}\,V(p, \theta_1)\,V(- p, \theta_2)\,,
\end{eqnarray}

\noindent
where the $1/2$ comes from the second order of the perturbative
expansion. After standard rearrangements one gets

\begin{eqnarray}
\label{III-51a}
\Gamma_{VV;3{\rm b}}^{(1)}(p)\,&=&\,g^2\,\int \frac{d^4k}{(2 \pi)^4}\,d^4\theta \,{\cal V}_3(k - p, p, -k)\,{\cal V}_3(k, -p, -k + p)\,\left[-\,\frac{1}{k^2 (k + p)^2}\right]\nonumber\\
&\times&\,\left[ - 2 \,V(p, \theta)\,\left( k^2\,+\, {\not\! k}_{\alpha {\dot \alpha}}\,{\overline D}^{{\dot \alpha}}\,D^{\alpha}\right)V(-p, \theta)\right]\,+\,LDT\,,
\end{eqnarray}

\begin{eqnarray}
\label{III-51b}
\Gamma_{VV;3{\rm c}}^{(1)}(p)\,&=&\,g^2\,\int \frac{d^4k}{(2 \pi)^4}\,d^4\theta \,{\cal V}_3(k - p, p, -k)\,{\cal V}_3(-k+p, -p, k)\,\left[-\,\frac{1}{k^2 (k + p)^2}\right]\nonumber\\
&\times&\,\left[ - 2 \,k^2\,V(p, \theta)\,V(-p, \theta)\right]\,.
\end{eqnarray}

\noindent
Here, $LDT$ is short for all terms which are at the most logarithmically divergent. Furthermore, from Eq. (\ref{mlett:aII-30}) 

\begin{eqnarray}
\label{III-52}
&&{\cal V}_3(k - p, p, -k)\,{\cal V}_3(k, -p, -k + p)\nonumber\\
&=&\,-\,{\cal V}_3(k - p, p, -k)\,{\cal V}_3(-k+p, -p, k)\,=\,-\,\sin^2(k \wedge p)\,.
\end{eqnarray}

\noindent
As a result, the terms proportional to $k^2$, in the second brackets
in the right hand sides of Eqs. (\ref{III-51a}) and (\ref{III-51b}),
drop out in the sum $\Gamma_{VV;3{\rm
b}}^{(1)}(p)\,+\,\Gamma_{VV;3{\rm c}}^{(1)}(p)$. On the other hand,
the term proportional to ${\not\! k}_{\alpha {\dot \alpha}}$ in
Eq. (\ref{III-51a}) survives. From power counting follows that such
term might give rise to (dangerous) linear divergences. To see whether
this really happens, we start by expanding

\begin{equation}
\label{III-53}
-\,\frac{1}{k^2 (k + p)^2}\,,
\end{equation}

\noindent
around $p = 0$. It is then obvious that the would be linearly
divergent integral

\begin{equation}
\label{III-54}
\int \frac{d^4k}{( 2 \pi)^4}\, {\not\! k}_{\alpha {\dot \alpha}}\,\frac{1}{k^4}\,\sin^2(k \wedge p)\,,
\end{equation}

\noindent
vanishes by symmetric integration. As stated above, the even parity of
the trigonometric factor in Eq. (\ref{III-4}), eliminates the linear UV
divergences and also the linear UV/IR infrared
divergences. To summarize:

\begin{equation}
\label{III-55}
\Gamma_{VV;3{\rm b}}^{(1)}(p)\,+\,\Gamma_{VV;3{\rm c}}^{(1)}(p)\,=\,LDT\,.
\end{equation}

We turn next into computing the ghost contributions to
$\Gamma^{(1)}_{VV}$. A direct consequence of the $D$-algebra is that
graphs containing any of the vertices $\Gamma^{(0)}_{c^{\prime} V c}$,
$\Gamma^{(0)}_{{\overline c}^{\prime} V {\overline c}}$,
$\Gamma^{(0)}_{c^{\prime} V V c}$, or $\Gamma^{(0)}_{{\overline
c}^{\prime} V V {\overline c}}$, depicted in Fig. \ref{vertices}, only
contribute $LDT$. We shall therefore concentrate on the diagrams which
might provide quadratic and/or linear divergent contributions to
$\Gamma^{(1)}_{VV}$. These are the graphs (d) and (e) of
Fig. \ref{twopoint}.

The calculation of the tadpole contributions (graphs (d) and (e) in
Fig.  \ref{twopoint}) $\Gamma _{VV;3{\rm d}}^{\left(1\right)}\left(p\right)$
is straightforward and yields

\begin{equation}
\label{III-6}
\Gamma _{VV;3{\rm d}}^{\left(1\right)}\left(p\right)  =  +\frac{g^{2}}{3}\int \frac{d^{4}k}{\left(2\pi \right)^{4}}\,d^{4}\theta \,\frac{\mathcal{V}_{4}^{(1)}
\left(k,p,-p,-k\right)}{k^{2}}\,V\left(p,\theta \right)\,V\left(-p,\theta 
\right)\,.
\end{equation}

\noindent
The same expression arises for $\Gamma _{VV;3{\rm e}}^{\left(1\right)}\left(p\right)$. Then, after using (\ref{III-2}), one obtains

\begin{equation}
\label{III-7}
\Gamma _{VV;3{\rm d}}^{\left(1\right)}\left(p\right)\,+\, \Gamma _{VV;3{\rm e}}^{\left(1\right)}\left(p\right) \,=\,\frac{4}{3}\,g^2\,A\,.
\end{equation}

The evaluation of the ghost loop contributions (graphs (f) and (g) in
Fig. \ref{twopoint}) is a little bit more involved. By applying the
Feynman rules we obtain

\begin{eqnarray}
\label{III-8}
&&\Gamma _{VV;3{\rm f}}^{\left(1\right)}\left(p\right)\, = \, \left(-1\right)\times 2\times \frac{1}{2!}\times g^{2}\nonumber\\
&\times& \int \frac{d^{4}k}{\left(2\pi \right)^{4}}d^{4}\theta _{1}d^{4}\theta _{2}\,\mathcal{V}_3\left(k,p,-p-k\right){\cal V}_3\left(p+k,-p,-k\right)\,V\left(p,\theta _{1}\right)\,V\left(-p,\theta _{2}\right)\nonumber\\
&\times&   \left[i\,\left(\overline{D}^{2}D^{2}\right)\frac{\delta _{12 }}{\left(k+p\right)^{2}}\right]\left[i\,\left(D^{2}\overline{D}^{2}\right)\frac{\delta _{12 }}{k^{2}}\right]\,, 
\end{eqnarray}

\noindent
where the $-1$ arises from the ghost loop and

\begin{equation}
\label{III-9}
\mathcal{V}_3\left(k,p,-p-k\right)\,{\cal V}_3\left(p+k,-p,-k\right)\,=\,-\,\sin^2(k \wedge p)\,.
\end{equation}

\noindent
It turns out that $\Gamma _{VV;3{\rm f}}^{\left(1\right)}\,=\,\Gamma
_{VV;3{\rm g}}^{\left(1\right)}$. Therefore,

\begin{equation}
\label{III-10}
\Gamma _{VV;3{\rm f}}^{\left(1\right)}\left(p\right)\,+\,\Gamma _{VV;3{\rm g}}^{\left(1\right)}\left(p\right)\,=\,-\,2\,g^2\,A\,+\,LDT\,.
\end{equation}

\noindent
We stress, once again, that the would be linear divergences in
Eqs. (\ref{III-6}) and (\ref{III-10}) are wiped out by symmetric
integration.

From Eqs. (\ref{III-3}), (\ref{III-55}), (\ref{III-7}), and
(\ref{III-10}) follows that the quadratic UV and the UV/IR infrared
divergences do not show up for ${\cal N} = 1$, in any arbitrary
covariant gauge.

We shall next investigate the consequences of adding one matter
superfield to get the ${\cal N} = 2$ theory. The amplitudes associated
with the graphs (h) and (i) in Fig. \ref{twopoint} are

\begin{equation}
\label{III-11}
\Gamma_{VV;3{\rm h}}^{(1)}(p) =  2\,( i g^{2})\,\int \frac{d^{4}k}{\left(2\pi \right)^{4}}d^{4}\theta\, \mathcal{V}_4\left(k,p,-p,-k\right)\,\left[ i\,\left(\overline{D}^{2}D^{2}\right)\frac{\delta _{11}}{k^{2}}\right]\, V\left(-p,\theta \right)V\left(p,\theta \right)\,
\end{equation}

\noindent
and

\begin{eqnarray}
\label{III-12}
\Gamma_{VV;3{\rm i}}^{(1)}(p) \, &=& \, (-\,2\, g)^2\,\int \frac{d^{4}k}{\left(2\pi \right)^{4}}d^{4}\theta _{1}d^{4}\theta _{2}\,\mathcal{V}_3\,\left(-p-k,p,k\right)\,\mathcal{V}_3\left(-k,-p,p+k\right)
\nonumber \\
&\times  & \left[i\,\left(\overline{D}^{2}D^{2}\right)\frac{\delta _{12}}{\left(k+p\right)^{2}}\right]\left[i\,\left(D^{2} \overline{D}^{2}\right)\frac{\delta _{12}}{k^{2}}\right]\,V\left(p,\theta _{1}\right)\,V\left(-p,\theta _{2}\right)\,.
\end{eqnarray}

\noindent
By taking into account Eqs. (\ref{III-2}) and (\ref{III-9}) one obtains

\begin{equation}
\label{III-13}
\Gamma_{VV;3{\rm h}}^{(1)}(p) \,=\,-\,4\,g^2\,A\,,
\end{equation}

\noindent
and

\begin{equation}
\label{III-14}
\Gamma_{VV;3{\rm i}}^{(1)}(p) \,=\,4\,g^2\,A\,+\,LDT\,.
\end{equation}

\noindent
Therefore, up to LDT, $\Gamma_{VV;3{\rm h}}^{(1)}\,+\,\Gamma_{VV;3{\rm
i}}^{(1)}\,=\,0$ implying in the absence of quadratic UV and UV/IR
infrared divergences in the matter sector and, therefore, in the full
${\cal N} = 2$ NCSQED$_4$. The validity of this conclusion for ${\cal
N} = 4$ is clear. Furthermore the UV logarithmic divergences are also
absent in ${\cal N} =4$ in agreement with \cite{Za1,Zanon}.

\section{One loop-contributions to the vector gauge superfield three-point function ${\Gamma}_{VVV}^{(1)}$}
\label{sec:level4}

We present in this Section the computation of the one-loop corrections
to the V gauge field three-point function in the covariant superfield
formalism. The divergence structure of the superfield formulation is,
as we shall see, substantially different of that encountered by
Matusis et al. \cite{Mat} in the component formulation. As for the
background field three-point function computed in \cite{Za1,Zanon} our
results play an essential role when considering insertions in higher
order corrections such as the one indicated in
Fig. \ref{higherorder}.

The one-loop diagrams contributing to the the three-point gauge field
function $\Gamma^{(1)}_{VVV}$ contain, generally speaking, a planar
and a nonplanar part. The planar parts will exhibit, at the most,
quadratic UV divergences, in agreement with Eq. (\ref{II-32}). These
divergences will be eliminated by dimensional regularization. The
linear UV divergences are always wiped out by symmetric
integration. Renormalization takes care of the logarithmic UV
divergences. As for the nonplanar parts the situation will be seen to
be more involved. Due to the peculiar structure of the Moyal
trigonometric factors, quadratic UV divergences do not translate into
quadratic UV/IR infrared singularities, but rather into linear and
logarithmic ones. Hereafter, we shall refer to this effect as to the
softening mechanism of divergences. There are, of course,
linear infrared divergences arising from the would be linear UV
divergences through the UV/IR mechanism. Finally, the logarithmic
UV/IR infrared singularities are harmless and shall be left out of
consideration.

Before facing the problem of selecting the diagrams of interest, we
found appropriate to exemplify how the softening mechanism of
divergences works. To this end, let us first consider the integral

\begin{equation}
\label{IV-1}
I^{\mu}(p_1, p_2, p_3)\,\equiv\,-\frac{1}{4}\,\int \,\frac{d^4k}{(2 \pi)^4}\,\left[\sin \left(2k \wedge p_1 \right)\,+\,\sin \left(2k \wedge p_2 \right)\,+\,\sin \left(2k \wedge p_3 \right)\right]\,\frac{k^{\mu}}{k^4}\,.
\end{equation}

\noindent
An straightforward computation yields

\begin{equation}
\label{IV-2}
I^{\mu}\,(p_1, p_2, p_3)\,\xrightarrow[p_1, p_2, p_3 \rightarrow 0]{}\,\frac{i}{2 \pi^2}\,\Theta^{\mu \nu}\,\left( \frac{p_{1 \nu}}{p_1 \circ p_1}\,+\,\frac{p_{2 \nu}}{p_2 \circ p_2}\,+\,\frac{p_{3 \nu}}{p_3 \circ p_3}\right)\,,
\end{equation}

\noindent
where

\begin{equation}
\label{IV-3}
p \circ p\,\equiv\,p^{\mu}\,\left( \Theta^2 \right)_{\mu \nu}\,p^{\nu}\,.
\end{equation} 

\noindent
From observation follows that $I^{\mu}\,(p_1, p_2, p_3)$ exhibits a linear infrared divergence. However, a nonplanar Feynman diagram whose corresponding amplitude is proportional to 

\begin{equation}
\label{IV-4}
\sin(p_1 \wedge p_2)\, I^{\mu}\,(p_1, p_2, p_3)\,,
\end{equation}

\noindent
will be finite if only one of the momenta goes to zero and vanishing if one lets all momenta to zero simultaneously. On the other hand, an amplitude proportional to

\begin{equation}
\label{IV-5}
\cos(p_1 \wedge p_2)\, I^{\mu}\,(p_1, p_2, p_3)\,,
\end{equation}

\noindent
will certainly have a linear divergence at $p_i \rightarrow 0$. Needless to say, the conversion of quadratic UV divergences into linear UV/IR infrared divergences is also possible through this softening mechanism of divergences.

We turn back into our main line of development and look for the diagrams which can make IR harmful contributions to $\Gamma^{(1)}_{VVV}$. To this end, one is to take into consideration the $D$-algebra, the parity of the Feynman integrands, and the softening mechanism described above. 

In this way we have found that the potentially harmful one-loop diagrams contributing to $\Gamma^{(1)}_{VVV}$ are those depicted in Figs. \ref{threepoint1}, \ref{threepoint2}, \ref{threepoint3}, \ref{threepoint4}, \ref{threepoint5}, \ref{threepoint6}, and \ref{threepoint7}. To systematize our presentation as well as to facilitate the verification of our calculations, we shall write all the three-point amplitudes $\Gamma_{VVV}^{(1)}$ as follows

\begin{equation}
\label{IV-5a} 
\Gamma_{VVV}^{(1)}\,=\,\left[\frac{1}{n!}\right]\,\times\,[t]\,\times\,[v]\,\times\,\int\,\left[\dk \, d\theta\right]\,\times\,\left[F_T\right]\,\times\,[P]\,\times\,{\mathbb D}_{\theta}\,+\,permutations\,.
\end{equation}

\noindent
Here, $1/n!$ comes from the order of the perturbative expansion, $t$
is the topological factor, $v$ is the numerical factor associated with
the vertices, $d\theta$ is the fermionic measure, $F_T$ is the
trigonometric factor provided by the noncommutativity, $P$ is the
product of $\theta$-independent factors in the propagators, ${\mathbb
D}_{\theta}$ is the $\theta$-dependent part of the integrand, and one
is to sum over the appropriate permutations of the external momenta.

For the tadpole graph in Fig. \ref{threepoint1} one has that $n = t = 1$, $v = - g^3/36$,

\begin{eqnarray}
\label{IV-6}
F_{T;5}\,&=&\,
\mathcal{V}^{(1)}_5\left(k,-k,p_{1},p_{2},p_{3}\right)\nonumber\\
&  = &\, 3\,\cos \left[p_{1}\wedge \left(p_{3}-k\right)+p_{3}\wedge k\right]\sin \left(k\wedge p_{2}\right)\nonumber \\
 & + & \,3\cos \left[p_{2}\wedge \left(p_{3}-k\right)+p_{3}\wedge k\right]\sin \left(k\wedge p_{1}\right)\nonumber \\
 & + & \,3\cos \left[p_{2}\wedge \left(p_{1}-k\right)+p_{1}\wedge k\right]\sin \left(k\wedge p_{3}\right)\,,
\end{eqnarray}

\noindent
as can be seen from Eq. (\ref{mlett:dII-30}), $P = -i/k^2$, and

\begin{eqnarray}
\label{IV-7}
{\mathbb D}_{\theta}\,&=&\,\delta_{12}\,\left[ \overline{D}_1^{2}D^{\alpha}_1 D_{2 \alpha}\,\left(1 - (1 - a)\,\frac{1}{k^{2}}\{D_1^2 , {\overline D}^2_1\}\right)\delta _{12}\right]
V\left(p_{1},\theta_1 \right) V\left(p_{2},\theta_1 \right) V\left(p_{3},\theta_1 \right)\nonumber\\
&=&\,\delta_{12}\,\left[ \overline{D}_1^{2}D^{\alpha}_1 D_{2 \alpha}\,\delta _{12}\right]
V\left(p_{1},\theta_1 \right) V\left(p_{2},\theta_1 \right) V\left(p_{3},\theta_1 \right)\,.
\end{eqnarray}

\noindent
As in the case of the two-point function the term in the propagator proportional to $(1 - a)$ drops out. By putting all these together one ends up with, 

\begin{eqnarray}
\label{IV-8}
&&{\Gamma} _{VVV;5}^{\left(1\right)}\left(p_{1},p_{2},p_{3}\right)\nonumber\\
&&  = \, -\frac{i g^{3}}{18}\,\int \frac{d^{4}k}{\left(2\pi \right)^{4}}d^{4}\theta \,\frac{F_{T;5}}{k^{2}}\,V\left(p_{1},\theta \right)V\left(p_{2},\theta 
\right)V\left(p_{3},\theta \right)\,+\,AP\,,
\end{eqnarray}

\noindent
where $AP$ means that one is to sum over all permutations of the
external momenta. By power counting Eq. (\ref{IV-8}) is quadratically
UV divergent but, on the other hand, Eq. (\ref{IV-6}) tell us that the
planar part vanishes, implying in the absence of UV
divergences. Hence, what we have to investigate are the consequences
of the UV/IR mechanism fully contained in the nonplanar part. A direct
calculation shows that

\begin{eqnarray}
\label{IV-9}
{\Gamma}_{VVV;5}^{\left(1\right)}\left(p_{1},p_{2},p_{3}\right) \,& 
= &\, -\, \frac{i}{8 \pi^2}\,\left\{ \sin \left(p_{1}\wedge p_{3}\right)
\left[\frac{1}
{p_{3}\circ p_{3}}-\frac{1}{p_{1}\circ p_{1}}\right]\right.\nonumber\\
 && + \sin \left(p_{2}\wedge p_{3}\right)\left[\frac{1}{p_{3}\circ p_{3}}-\frac{1}{p_{2}\circ p_{2}}\right]\nonumber\\
 && \left.  +  \sin \left(p_{2}\wedge p_{1}\right)\left[\frac{1}{p_{1}\circ p_{1}}-\frac{1}{p_{2}\circ p_{2}}\right]\right\}\,B\,+\,AP\,.
\end{eqnarray} 

\noindent
For arriving at Eq. (\ref{IV-9}) we have used

\begin{subequations}
\label{IV-10}
\begin{eqnarray}
&&\int \frac{d^{4}k}{\left(2\pi \right)^{4}}\frac{\sin \left(2k\wedge p\right)}{k^{2}} \,=\,  0\,,\label{mlett:aIV-11}\\
&&\int \frac{d^{4}k}{\left(2\pi \right)^{4}}\frac{\cos \left(2k\wedge p\right)}{k^{2}}  =  \frac{1}{4\pi ^{2}p\circ p}\label{mlett:bIV-11}\,,
\end{eqnarray}
\end{subequations}

\noindent
and

\begin{equation}
\label{IV-11}
B\,\equiv\,g^{3}\int d^{4}\theta \,V\left(p_{1},\theta \right)V\left(p_{2},\theta \right)V\left(p_{3},\theta \right)\,.
\end{equation}

\noindent
It is easy to verify that momentum conservation enforces 

\begin{equation}
\label{IV-12}
{\Gamma} _{VVV;5}^{\left(1\right)}\left(p_{1},p_{2},p_{3}\right) \, = \,0\,,
\end{equation}

\noindent
implying in the absence of UV/IR infrared divergences as well. One can
convince oneself that the trigonometric factor corresponding to the
tadpoles involving the vertices $\Gamma_{(\overline{D}^2
V)(DV)(DV)V}^{(0)}$ and $\Gamma_{(\overline{D}DV)({\overline
D}V)(DV)V}^{(0)}$ (see Fig. \ref{vertices}) are proportional to
$\sin(p_1 \wedge p_2)$ and, therefore, the would be linear UV/IR
infrared divergence is softened and becomes harmless.

The diagrams in Fig. \ref{threepoint2} have in common the four-point
vertex $\Gamma_{(\overline{D}^2DV)(DV)VV}^{(0)}$. We focus first on
diagram (a). We have that $n = 2$, $t = 4$, $v = -ig^3/12$, and

\begin{equation}
\label{IV-13}
P\,=\,\frac{(-i)^2}{k^2\,(k - p_3)^2}\,.
\end{equation}

\noindent
For the trigonometric factors an straightforward calculation yields

\begin{equation}
\label{IV-14}
F_{T;6{\rm a}}\,=\,-2\,\cos(p_1 \wedge p_2)\,F_T^{odd}\,+\,2\,\sin(p_1 \wedge p_2)\,F_{T;6{\rm a}}^{even}\,,
\end{equation}

\noindent
where

\begin{subequations}
\label{IV-15}
\begin{eqnarray}
&& F_T^{odd}\,=\,-\,\frac{1}{4}\,\left[ \sin \left(2 k \wedge p_1 \right)\,+\,\sin \left(2 k \wedge p_2 \right)\,+\,\sin \left(2 k \wedge p_3 \right)\right]\,,\label{mlett:aIV-15}\\
&& F_{T;6{\rm a}}^{even}\,=\,-\,\frac{1}{4}\,\left[ \cos \left(2 k \wedge p_1 \right)\,-\,\cos \left(2 k \wedge p_2 \right)\right]\,.\label{mlett:bIV-15}
\end{eqnarray}
\end{subequations}

\noindent
As for the $\theta$ dependent factors, we obtain,

\begin{eqnarray}
\label{IV-16}
{\mathbb D}_{\theta ; 6{\rm a}}\,&=&\, -\,2\,\left[ (k - p_3)^2\,V(p_3, \theta_2)\,+\,({\not\! k} - {\not\! p_3})_{\alpha {\dot \alpha}}\,\left({\overline D}^{{\dot \alpha}} D^{\alpha} V(p_3, 
\theta_2 \right)\,+\,\cdots\right]\,\nonumber\\
&\times&\,\delta_{12}\,V(p_1, \theta_1)\,V(p_2, \theta_1)\,,
\end{eqnarray}

\noindent
where we omitted all terms leading to contributions which are at the most logarithmically divergent. Observe that the term proportional to $(1 -a)$ in the $V$-propagator does not contribute again. 

Power counting says that the diagram (a) is UV quadratically divergent
although, as before, the corresponding trigonometric factor does not
contain a planar part. The amplitude is, then, UV finite and we
concentrate on studying the outcomes of the UV/IR mechanism. After
expanding Eq. (\ref{IV-13}) around $p_3 = 0$, the expression for the
amplitude associated with the graph (a) can be cast

\begin{eqnarray}
\label{IV-17}
\Gamma_{VVV;6{\rm a}}^{(1)}\,&=&\,\frac{ig^3}{6}\,\int \dk\,d\theta\,F_{T;6{\rm a}}\,\left[ \frac{1}{(k^2)^2}\,+\,2\,p^{\mu}_3\,\frac{k_{\mu}}{(k^2)^3}\,+\,\cdots \right]
\nonumber\\
\,&\times&\,\left[ (k - p_3)^2\,V(p_3, \theta)\,+\,({\not\! k} - {\not\! p_3})_{\alpha {\dot \alpha}}\,\left({\overline D}^{{\dot \alpha}} D^{\alpha} V(p_3, \theta )\right)\,+\,\cdots\right]
\nonumber\\
\,&\times&\,V(p_1, \theta)\,V(p_2, \theta)\,+\,CP\,,
\end{eqnarray}

\noindent
where $CP$ means sum over cyclic permutations of the external
momenta. By collecting terms of equal power in $k$, we write

\begin{equation}
\label{IV-18}
\Gamma_{VVV;6{\rm a}}^{(1)}\,\equiv\,\gamma_{6{\rm a}}^{[2]}\,+\,\gamma_{6
{\rm a}}^{[1]}\,+\,LDT\,,
\end{equation}

\noindent
where 

\begin{equation}
\label{IV-19}
\gamma_{6{\rm a}}^{[2]}\,=\,\left(\frac{ig^3}{6}\right)\,2\,\sin(p_1 \wedge p_2)\, \int \dk\,d\theta\,F_{T;6{\rm a}}^{even}\frac{1}{k^2}\,V(p_1, \theta)V(p_2, \theta)V(p_3, \theta)\,+\,CP\,,
\end{equation}

\begin{eqnarray}
\label{IV-20}
\gamma_{6{\rm a}}^{[1]}\,&=&\,-\,\left(\frac{ig^3}{6}\right)\,2\,\cos(p_1 \wedge p_2)\, \int \dk\,d\theta\,F_{T}^{odd}\frac{1}{\left(k^2\right)^2}
\nonumber\\
\,&\times&\,{\not\! k}_{\alpha {\dot \alpha}}\left({\overline D}^{{\dot \alpha}} D^{\alpha} V(p_3, \theta )\right)\,V(p_1, \theta)V(p_2, \theta)\,+\,CP\,,
\end{eqnarray}

\noindent
and the superscript makes reference to the power of $k$.

A word of caution is here in order. Two terms of the form

\begin{equation}
\label{IV-20a}
p_3^{\mu}\,\cos(p_1 \wedge p_2)\,\int\,\dk\,F_T^{odd}\frac{k_{\mu}}{\left(k^2\right)^2}\,B\,,
\end{equation}

\noindent
occur in the right hand side of Eq. (\ref{IV-20}). Expressions of this
type arise as a result of expanding the factor $P$ (see
Eq. (\ref{IV-5a})) around zero external momenta. In the present case
they cancel out between themselves, giving no contribution to linear
UV/IR infrared divergences. However, we would like to remark that
individually they also vanish,  since after performing the sum over the
permutations of the external momenta one finds

\begin{equation}
\label{IV-20b}
\left(p_1\,+\,p_2\,+\,p_3\right)^{\mu}\,\cos(p_1 \wedge p_2)\,
\int\,\dk\,F_T^{odd}\frac{k_{\mu}}{\left(k^2\right)^2}\,B\,,
\end{equation}

\noindent
which is set to zero by momentum conservation.

By carrying out the momentum integrals in Eqs. (\ref{IV-19}) and
(\ref{IV-20}) and after $D$-algebra rearrangements, one obtains, respectively,

\begin{equation}
\label{IV-21}
\gamma_{6 {\rm a}}^{[2]}\,=\,-\,\left(\frac{i}{12}\right)\,\sin(p_1 \wedge p_2)\,\frac{1}{4 \pi^2} \left( \frac{1}{p_1 \circ p_1}\,-\,\frac{1}{p_2 \circ p_2}\right) \, B \,+\,CP
\end{equation}

\noindent
and

\begin{equation}
\label{IV-22}
\gamma_{6{\rm a}}^{[1]}\,=\,4\,\left(\frac{i}{6}\right)\,\cos(p_1 \wedge p_2)\,I^{\mu}(p_1,p_2,p_3)
\,N_{\mu}(p_1,p_2,p_3)\,.
\end{equation}

\noindent
Here,

\begin{equation}
\label{IV-23}
N_{\mu}(p_1,p_2,p_3)\,\equiv g^3\,\left( \sigma_{\mu} \right)_{\alpha {\dot \alpha}} \int d\theta \left( D^{\alpha} V(p_1,\theta)\,{\overline D}^{{\dot \alpha}}V(p_3,\theta)\,V(p_2,\theta)\,+\,AP\right)\,,
\end{equation}

\noindent
while $I^{\mu}$ was already defined in Eq. (\ref{IV-1}).

After recalling momentum conservation one concludes that

\begin{equation}
\label{IV-24}
\gamma_{6{\rm a}}^{[2]}\,=\,0\,.
\end{equation}

\noindent
Thus we are left with a harmful linear UV/IR infrared divergence in
$\Gamma_{VVV;6{\rm a}}^{(1)}$ given at Eq. (\ref{IV-22}).

The diagrams (b), (c), and (d) in Fig. \ref{threepoint2} have in
common that the terms proportional to $(1 - a)$ drop out, as in the
case of diagram (a). Since $ F_{T;6{\rm b}}\,=\,-\,F_{T;6{\rm a}}$,
graph (b) also has no planar part. As for graphs (c) and (d) they  have a
logarithmically divergent planar part which demands UV
renormalization. For all of them, the nonplanar contribution
$\gamma^{[2]}$ is absent. We found that

\begin{equation}
\label{IV-25}
\sum_{j = 6{\rm b}}^{6{\rm d}}\,\gamma_{j}^{[1]}\,=\,3\,\left(\frac{i}{6}
\right)\,\cos(p_1 \wedge p_2)\,I^{\mu}(p_1,p_2,p_3)\,N_{\mu}(p_1,p_2,p_3)\,.
\end{equation}

On the other hand, the graphs (e), (f) and (g) turn out to be
proportional to $(1 - a)$. While (e) has no planar part, (f) and
(g) exhibit a logarithmic UV divergence. Again, $\gamma^{[2]} =
0$. We end up with

\begin{equation}
\label{IV-26}
\sum_{j = 6{\rm e}}^{6{\rm g}}\,\gamma_{j}^{[1]}\,=\,(1 - a)\,\left(\frac{ig^3}{6}\right)\,\cos(p_1 \wedge p_2)\,I^{\mu}(p_1,p_2,p_3)\,N_{\mu}(p_1,p_2,p_3)\,.
\end{equation}

Therefore, 

\begin{equation}
\label{IV-27}
\sum_{j = 6{\rm a}}^{6{\rm g}}\Gamma^{(1)}_{VVV;j}\,=\,(8 - a)\,
\left(\frac{i}{6}\right)\,\cos(p_1 \wedge p_2)\,I^{\mu}(p_1,p_2,p_3)\,
N_{\mu}(p_1,p_2,p_3)\,+\,LDT\,,
\end{equation}

\noindent
where the linear UV/IR infrared divergence, arising from the diagrams
in Fig. \ref{threepoint2}, is explicitly given.

We move next into computing the diagrams in
Fig. \ref{threepoint3}. Unlike those in Fig. \ref{threepoint2} they
involve the four-point vertex $\Gamma_{V(DV)(\overline{D}V)({\overline
D}DV)}^{(0)}$ (see Eq. (\ref{mlett:bII-2901})). An analysis quite
similar to that already presented enables one to conclude that the
planar part has, at  the most, a logarithmic UV divergence. A linear
UV/IR infrared divergence is present in each graph but, nevertheless,
cancels. That is

\begin{equation}
\label{IV-28}
\sum_{j = 7{\rm a}}^{7{\rm p}}\,\Gamma^{(1)}_{VVV;j}\,=\,LDT\,.
\end{equation}

The diagrams in Fig. \ref{threepoint4} involve the last four point
vertex $\Gamma_{V(DV)({\overline D}^2V)(DV)}^{(0)}$ quoted in
Eq. (\ref{mlett:cII-2901}). In the planar sector the situation is as in
the case of the diagrams in Fig. \ref{threepoint3}, only logarithmic UV
divergences show up. In the nonplanar sector the linear UV/IR
infrared divergences do not cancel and the final form for the
corresponding amplitude is

\begin{equation}
\label{IV-29}
\sum_{j = 8{\rm a}}^{8{\rm b}}\Gamma^{(1)}_{VVV;j}\,=\,-\,6\,a\,\left(
\frac{i}{6}\right)\,\cos(p_1 \wedge p_2)\,I^{\mu}(p_1,p_2,p_3)\,N_{\mu}(p_1,p_2,p_3)\,+\,LDT\,.
\end{equation}

We turn next into evaluating the graphs in Fig. \ref{threepoint5}. All
of them have in common, up to an overall sign, the trigonometric
factor

\begin{equation}
\label{IV-30}
F_{T;9}\,=\,\cos(p_1 \wedge p_2)\,F_T^{odd}\,+\,\sin(p_1 \wedge p_2)\,
F_{T;9}^{even}\,,
\end{equation}

\noindent
where $F_T^{odd}$ was defined in Eq. (\ref{mlett:aIV-15}) and

\begin{equation}
\label{IV-31}
F_{T;9}^{even}\,=\,-\,\frac{1}{4}\,\left[ 1\,-\,\cos(2k \wedge p_1)\,+\,
\cos(2k \wedge p_2)\,-\,\cos(2k \wedge p_3)\right]\,.
\end{equation}

\noindent
Hence, the planar part does not vanish. From the $D$-algebra follows
that quadratic UV divergences only arise in graphs (a) to (d) and
are taken care by dimensional regularization. For all graphs in
Fig. \ref{threepoint5}, linear UV divergences are killed by symmetric
integration, while the logarithmic ones are absorbed through
renormalization.

In principle there is nothing that could prevent the appearance of
quadratic UV/IR infrared divergences, from graphs (a) to (d), in
view of $F_{T;9}^{even}\, \neq \,0 $. Nevertheless, the presence of
$\sin(p_1 \wedge p_2)$ in Eq. (\ref{IV-30}) lowers the degree of this
divergence at least to linear. The softening mechanism, mentioned at
the beginning of this section, is again at work. For each graph, one
can verify that the UV/IR linear infrared divergences arising through
this mechanism are of the form

\begin{equation}
\label{IV-32}
\sin(p_1 \wedge p_2) \left( \frac{1}{p_1 \circ p_1}\,-\,\frac{1}{p_2 \circ p_2}\,+\,\frac{1}{p_3 \circ p_3}\right)\,,
\end{equation}

\noindent
and therefore cancel after symmetrizing in the external momenta.

On the other hand, for all the graphs, there are linear infrared
divergences which are the UV/IR counterparts of the would be linear UV
divergences. These divergences cancel when summing up over all graphs
in Fig. \ref{threepoint5}.

For the ghost graphs (a) and (b) in Fig. \ref{threepoint6} the
trigonometric factors are found to read

\begin{equation}
\label{IV-33}
F_{T;10{\rm a}}\,=\,F_{T;10{\rm b}}\,=\,-2\,\cos(p_1 \wedge p_2)\,F_T^{odd}\,+\,2\,\sin(p_1 \wedge p_2)\,F_{T;6{\rm a}}^{even}\,=\,F_{T;6{\rm a}}\,,
\end{equation}

\noindent
whereas for the others diagrams in Fig. \ref{threepoint6} one has

\begin{equation}
\label{IV-34}
F_{T;10^{\prime}}\,=\,-\,\cos(p_1 \wedge p_2)\,F_T^{odd}\,-\,\sin(p_1 \wedge p_2)\,F_{T;9}^{even}\,=\,-\,F_{T;9}\,.
\end{equation}

\noindent
The $D$-algebra signalizes again the presence of quadratic, linear and
logarithmic UV divergences in graphs (c) to (f) in Fig. \ref{threepoint6}, since their trigonometric
factor possesses a nonvanishing planar part. As it already happens in
connection with the graphs in Fig. \ref{threepoint5}, the linear UV/IR
infrared divergences arising from the softening mechanism vanish for
each graph. The remaining linear divergences do not cancel and we
obtain

\begin{eqnarray}
\label{IV-35}
&&\sum_{j = 10{\rm a}}^{10{\rm f}}\Gamma^{(1)}_{VVV;j}\,
=\,-\,4\,\left(\frac{i}{6}\right)\,\cos(p_1 \wedge p_2)\,I^{\mu}(p_1,p_2,p_3)\,N_{\mu}(p_1,p_2,p_3)\,+\,LDT\,.
\end{eqnarray}

To summarize, in NCSQED$_4$ and for ${\cal N}=1$, the one-loop
corrections to the three-point gauge superfield function are afflicted
by linear UV/IR infrared singularities. By collecting the calculations
presented in this section, Eqs. (\ref{IV-12}), (\ref{IV-27}),
(\ref{IV-28}), (\ref{IV-29}), and (\ref{IV-35}), we conclude that the
amplitude can be cast in the following form

\begin{equation}
\label{IV-36}
\Gamma^{(1)}_{VVV}=\,(4\,-\,7a)\,\left(\frac{i}{6}\right)\,\cos(p_1 
\wedge p_2)\,I^{\mu}(p_1,p_2,p_3)\,N_{\mu}(p_1,p_2,p_3)\,+\,LDT\,.
\end{equation}

\noindent
As can be seen, for the gauge $a = 4/7$ the three-point $V$ function
is free of linear UV/IR infrared divergences. To phrase it
differently, these divergences are localized in the gauge sector and are
a gauge artifact.

This result is not altered by the addition of one chiral matter
superfield (see Fig. \ref{threepoint7}). In fact, the contribution of
the tadpole graph (a) is proportional to that of the $V$-tadpole in
Eq. (\ref{IV-12}). Furthermore, the amplitudes corresponding to the
graphs (b) and (c) are proportional, respectively, to those of the
graphs (a) and (c) of Fig. \ref{threepoint6}. The linear UV/IR infrared
divergences resulting from the quadratic UV divergences, via the
softening mechanism, cancel out for each graph. As for the remaining
linear UV/IR infrared divergences, in diagrams (b) and (c), the
numerical coefficients are such as to secure their cancellation. The
generalization of these results to ${\cal N}=4$ is straightforward.

\section{Conclusions}
\label{sec:level5}

This work was dedicated to establish the consistency of NCSQED$_4$
within the covariant superfield formalism. As a first step, we
generalized the analysis of the two-point gauge field function
presented in \cite{bichl} by extending their results to an
arbitrary covariant gauge and for any matter content.

Our main contribution consists of a detailed study of the divergence
structure of the one-loop  three-point function of
the gauge superfield. The superfield formulation in an arbitrary
covariant gauge,  represents a significative
improvement with respect to the component field calculation presented
in \cite{Mat,RR1} for the same problem. At the very least, here
supersymmetry is kept operational at all stages of the
calculation. Unlike in the component formulation, we have found a
nonvanishing result for the linear UV/IR infrared divergences which
are, nevertheless, a gauge artifact. The
situation  resembles that encountered in QED$_4$ where the
infrared divergences disappear from the full two-point fermion Green
function in  a particular covariant gauge (Yennie's
gauge \cite{Yennie}).

The present work also plays a relevant role within the background
field formalism. Indeed, the computation of higher loop corrections to
the results encountered in \cite{Za1,Zanon} will necessarily
demand the insertion of the three-point $V$-function calculated in the
superfield covariant formalism (see Fig. \ref{higherorder}). Our
conclusion that the linear UV/IR infrared singularities are placed in
the gauge sector implies that higher order loop corrections to the
background field strength function will not be afflicted by harmful
UV/IR infrared singularities.

{\bf Acknowledgements.} This work was partially supported by Funda\c
c\~ao de Amparo \`a Pesquisa do Estado de S\~ao Paulo (FAPESP) and
Conselho Nacional de Desenvolvimento Cient\'\i fico e Tecnol\'ogico
(CNPq). H. O. G. and V. O. R. also acknowledge support from PRONEX
under contract CNPq 66.2002/1998-99. A. Yu. Petrov has been supported
by FAPESP, project No. 00/12671-7.

\newpage

\begin{figure}[ht]
\includegraphics{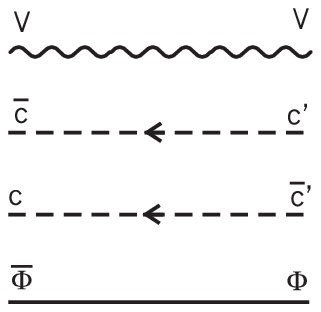}
\caption{\label{propagators} Free propagators. The arrow indicates the flux of ghost charge. }
\end{figure}

\begin{figure}[ht]
\includegraphics{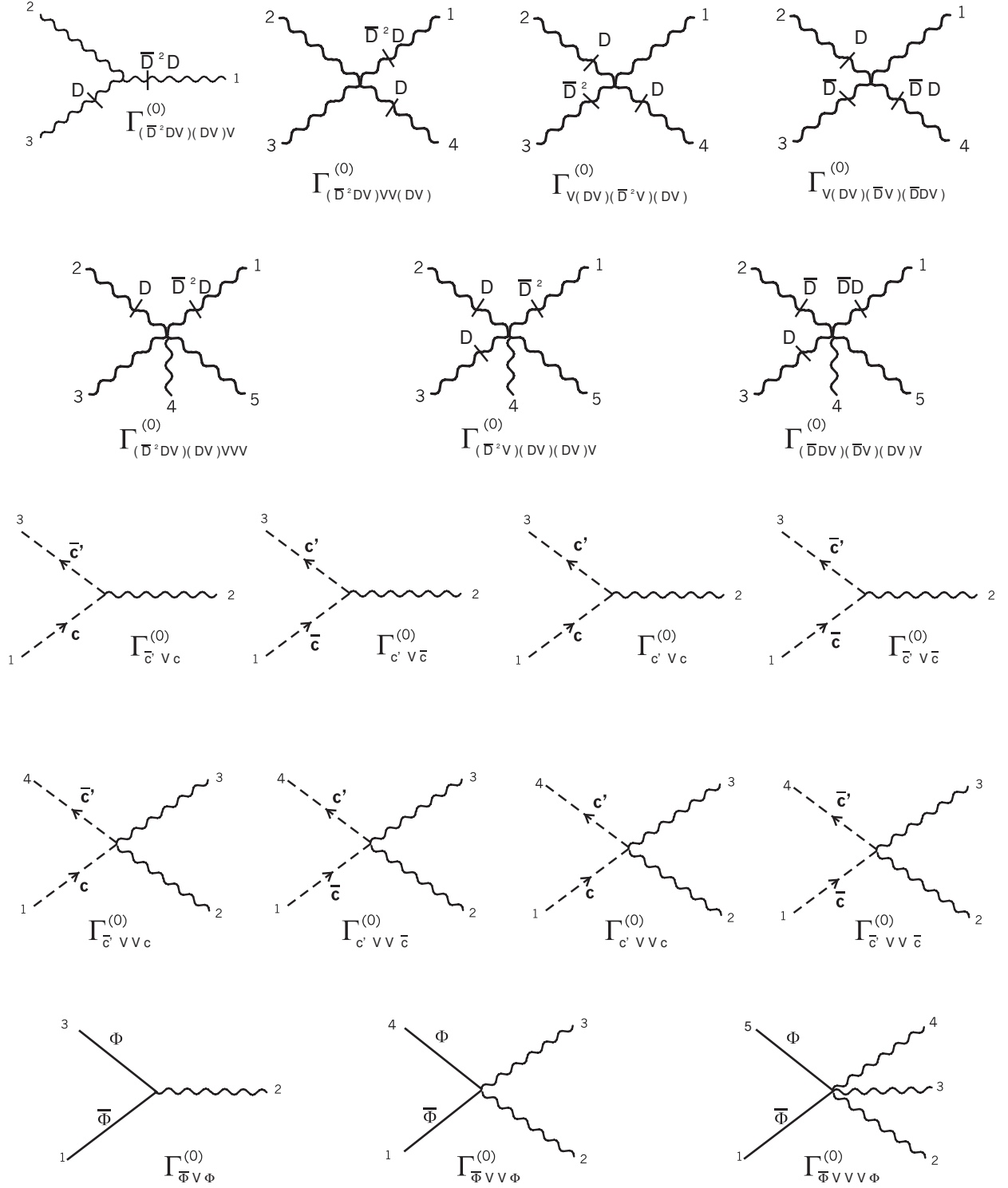}
\caption{\label{vertices}Elementary vertices.}  
\end{figure}

\begin{figure}[ht]
\includegraphics{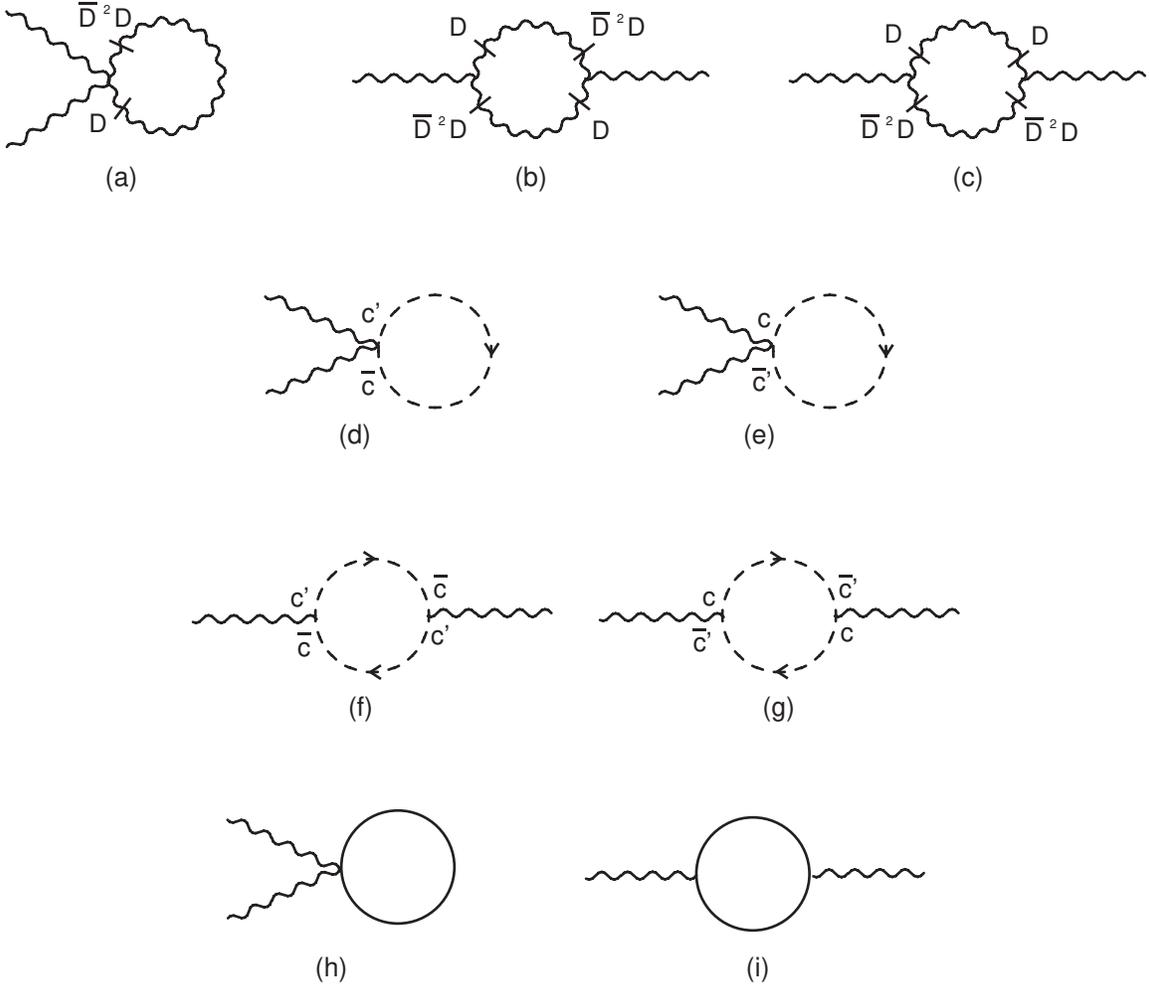}
\caption{\label{twopoint}Diagrams contributing to $\Gamma^{(1)}_{VV}$. }
\end{figure}

\begin{figure}[ht]
\includegraphics{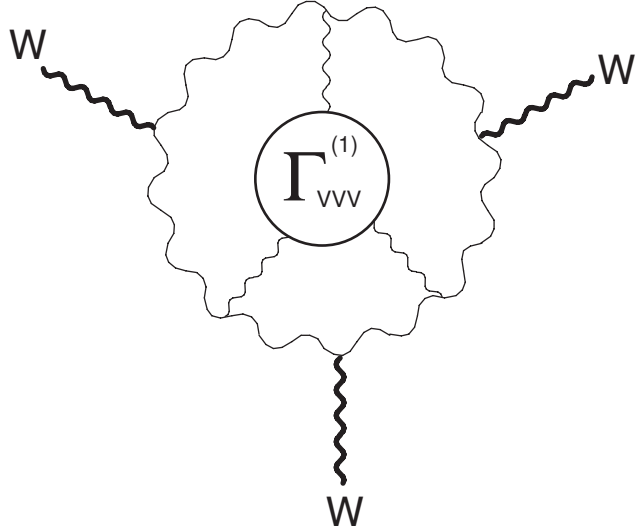}
\caption{\label{higherorder}An example of higher order correction to the three-point $W$ function including the one-loop three-point $V$ function.}  
\end{figure}

\begin{figure}[ht]
\includegraphics{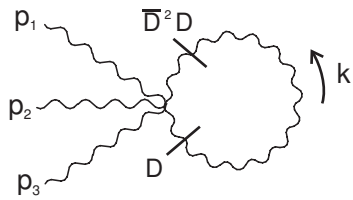}
\caption{\label{threepoint1} Tadpole contribution to $\Gamma^{(1)}_{VVV}$.}
\end{figure}

\begin{figure}[ht]
\includegraphics{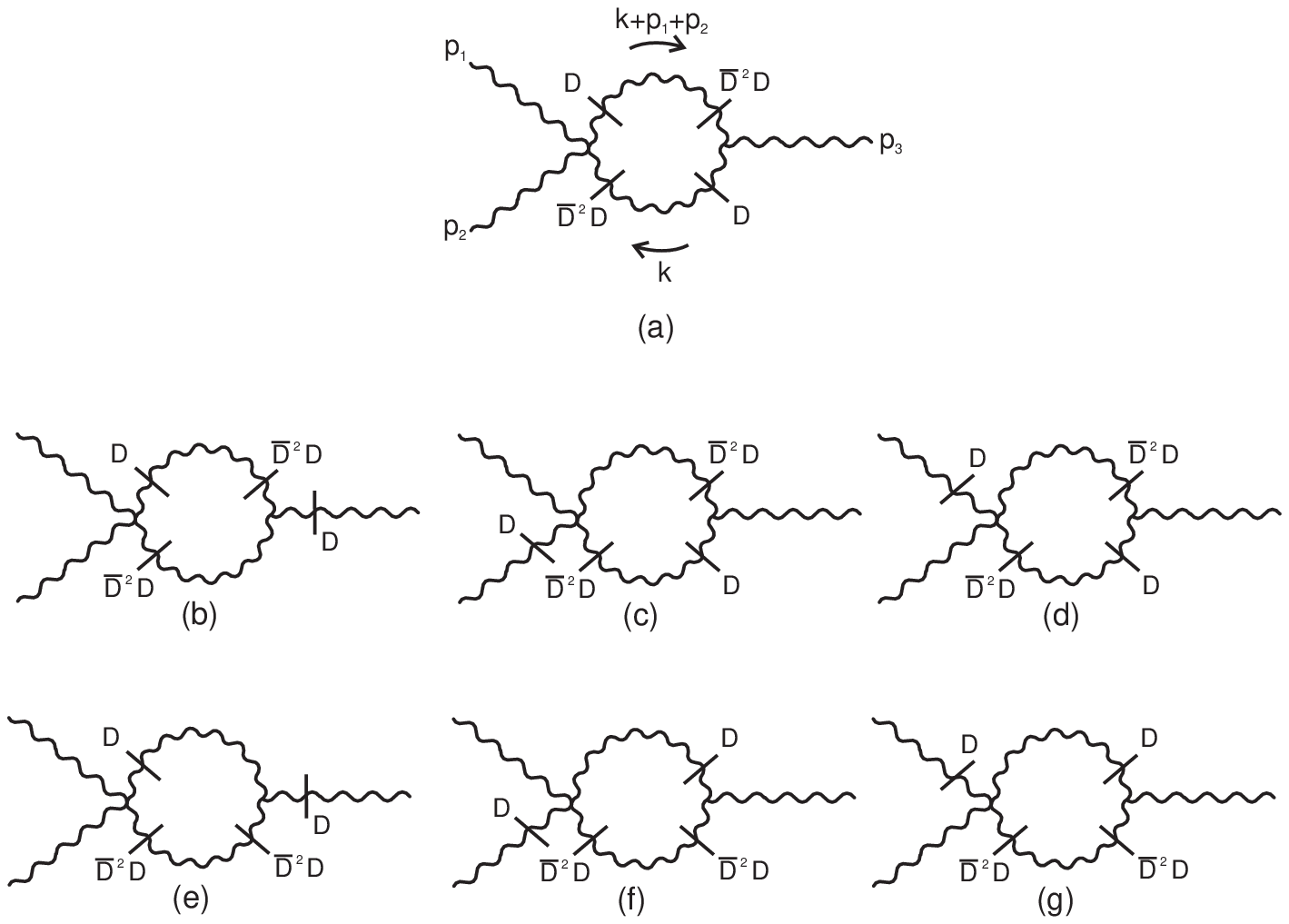}
\caption{\label{threepoint2} Contributions to $\Gamma^{(1)}_{VVV}$ involving the vertex $\Gamma_{(\overline{D}^2DV)(DV)VV}^{(0)}$.}
\end{figure}

\begin{figure}[ht]
\includegraphics{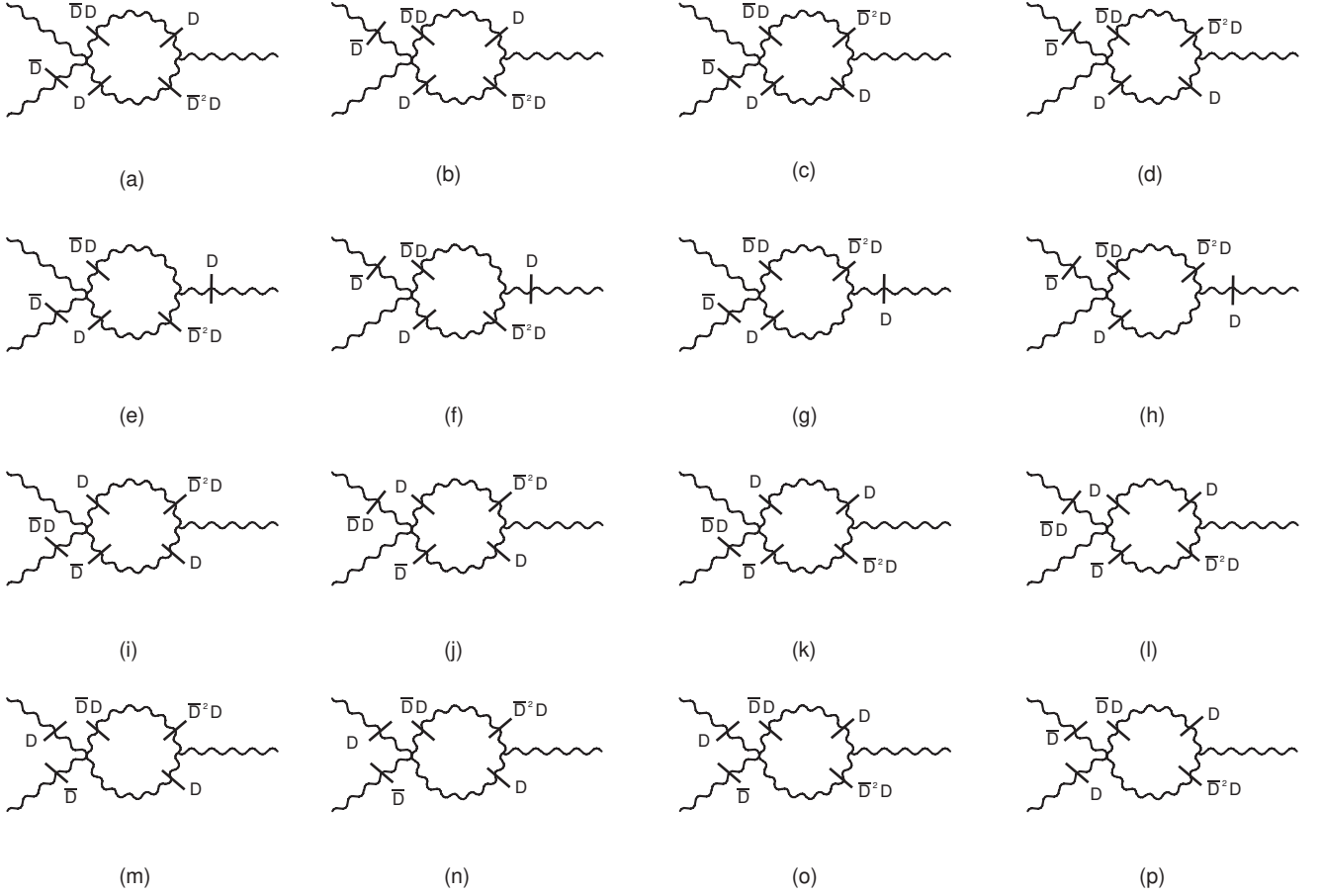}
\caption{\label{threepoint3} Contributions to $\Gamma^{(1)}_{VVV}$ involving 
the vertex $\Gamma_{V(DV)(\overline{D}V)({\overline D}DV)}^{(0)}$.}
\end{figure}

\begin{figure}[ht]
\includegraphics{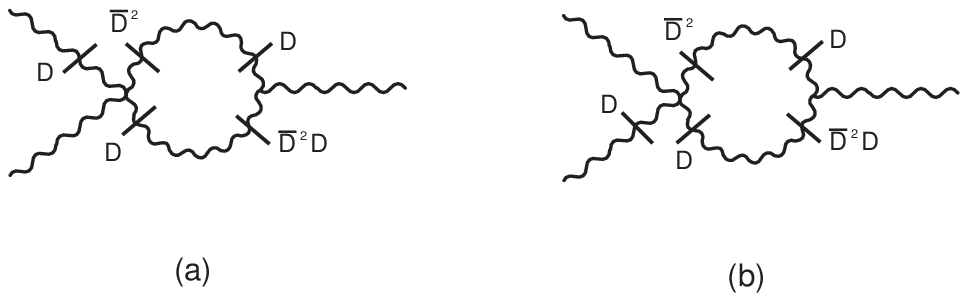}
\caption{\label{threepoint4} Contributions to $\Gamma^{(1)}_{VVV}$ involving the vertex $\Gamma_{V(DV)(\overline{D}^2V)(DV)}^{(0)}$.}
\end{figure}

\begin{figure}[ht]
\includegraphics{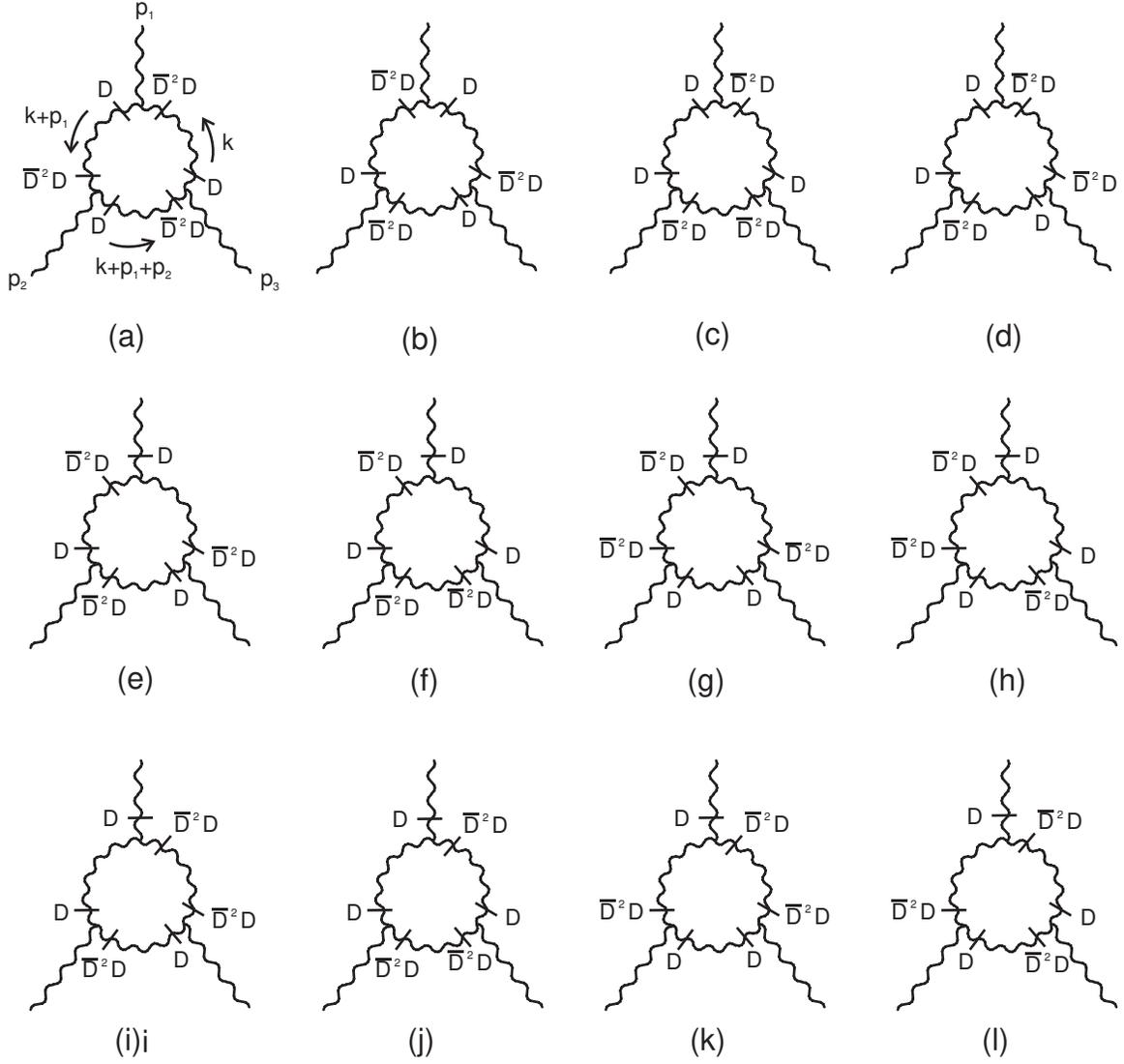}
\caption{\label{threepoint5} Contributions to $\Gamma^{(1)}_{VVV}$ involving the trilinear vertex only.}
\end{figure}

\begin{figure}[ht]
\includegraphics{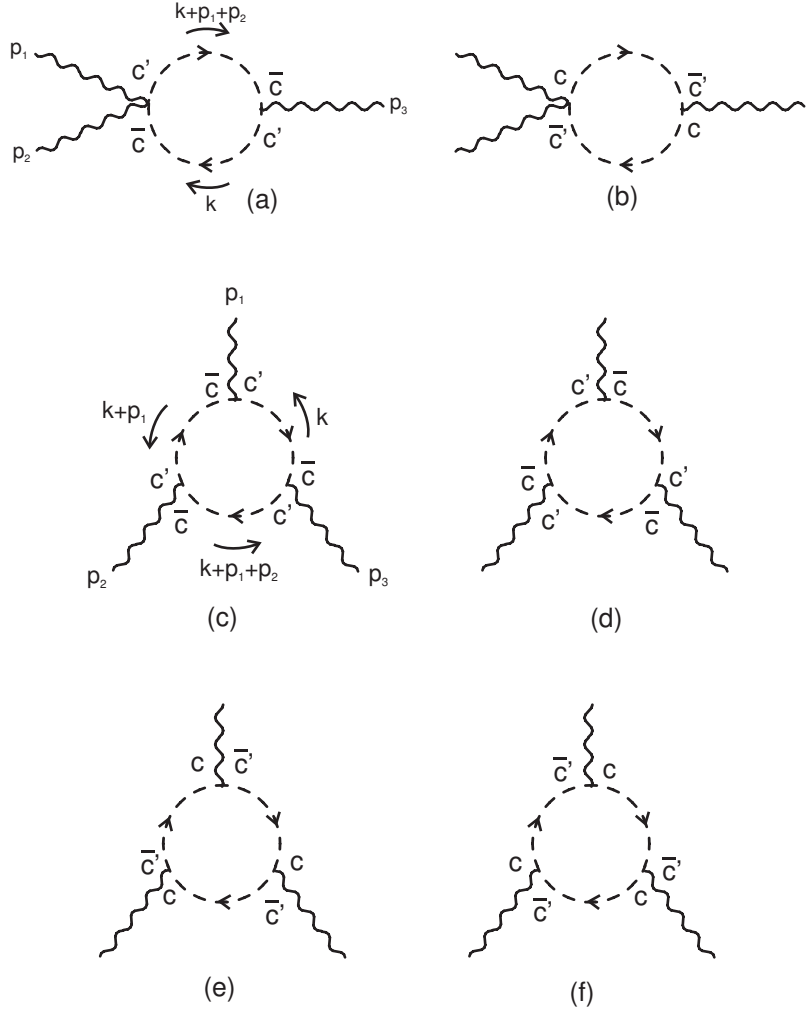}
\caption{\label{threepoint6} Ghost contributions to $\Gamma^{(1)}_{VVV}$.}
\end{figure}

\begin{figure}[ht]
\includegraphics{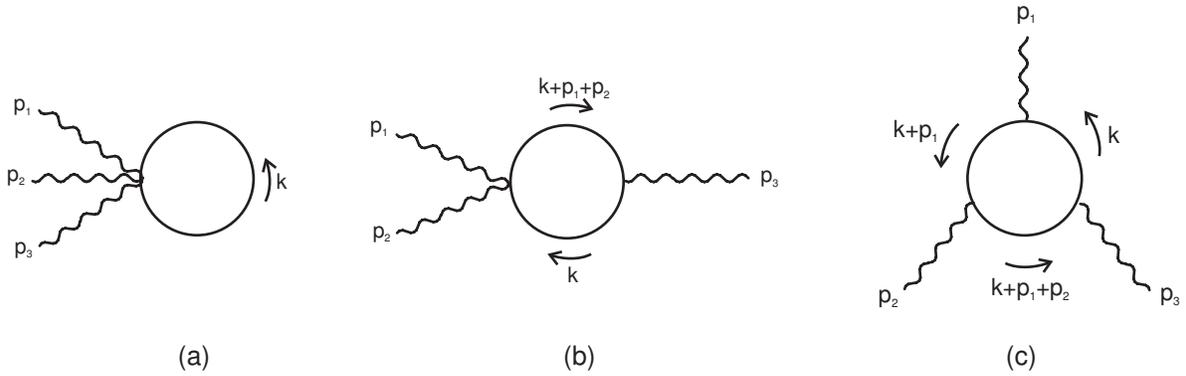}
\caption{\label{threepoint7} Matter contributions to $\Gamma^{(1)}_{VVV}$.}
\end{figure}


\begin{thebibliography}{99}
\bibitem {footnote1} To avoid confusion with the standard infrared
divergences, those originating through the UV/IR mechanism will be
referred to as UV/IR infrared divergences.
\bibitem {footnoteSGRS} In this paper, we shall follow the conventions
in Ref. \cite{SGRS}.
\end{thebibliography}
\end{document}